\documentclass[aps,pra,twocolumn,superscriptaddress]{revtex4-2}
\usepackage{times}
\usepackage{geometry} 		
\usepackage[pdftex]{graphicx}
\usepackage{amssymb,amsmath}
\usepackage{physics}
\usepackage{subfigure}
\usepackage{url}
\usepackage{braket}
\usepackage{framed}
\usepackage{enumerate}
\usepackage{autobreak}
\usepackage{hyperref}
\allowdisplaybreaks
\begin{document}
\title{
Anonymous estimation of intensity distribution of magnetic fields
with quantum sensing network}
\author{Hiroto Kasai}
\email{kasai-q@aist.go.jp}
\affiliation{
Graduate School of Pure and Applied Sciences, University of Tsukuba, 1-1-1 Tennodai, Tsukuba, Ibaraki 305-8571, Japan }
\affiliation{
Research Center for Emerging Computing Technologies, National institute of Advanced Industrial Science and Technology (AIST), Central2, 1-1-1 Umezono, Tsukuba, Ibaraki 305-8568, Japan }
\author{Yuki Takeuchi}
\email{yuki.takeuchi@ntt.com}
\affiliation{
NTT Communication Science Laboratories,
NTT Corporation, 3-1 Morinosato Wakamiya, Atsugi, 
Kanagawa 243-0198, Japan }
\author{Yuichiro Matsuzaki}
\email{matsuzaki.yuichiro@aist.go.jp}
\affiliation{
Research Center for Emerging Computing Technologies, National institute of Advanced Industrial Science and Technology (AIST), Central2, 1-1-1 Umezono, Tsukuba, Ibaraki 305-8568, Japan }
\author{Yasuhiro Tokura}
\email{tokura.yasuhiro.ft@u.tsukuba.ac.jp}
\affiliation{
Graduate School of Pure and Applied Sciences, University of Tsukuba, 1-1-1 Tennodai, Tsukuba, Ibaraki 305-8571, Japan }
\begin{abstract}
A quantum sensing network is used to simultaneously detect and measure physical quantities, such as magnetic fields, 
at different locations.
However, 
there is a risk that the measurement data is leaked to the third party during the communication. 
Many theoretical and experimental efforts have been made to realize a secure quantum sensing network where a high level of security is guaranteed.
In this paper, 
we propose a protocol to estimate statistical quantities of the target fields at different places without knowing individual value of the target fields.
We generate an enanglement between $L$ quantum sensors, 
let the quantum sensor interact with local fields, and perform specific measurements on them.
By calculating the quantum Fisher information
to estimate the individual value of the magnetic fields, 
we show that we cannot obtain any information of the value of the individual fields in the limit of large $L$.
On the other hand, in our protocol, 
we can estimate theoretically any moment of the field distribution by measuring a specific observable and 
evaluated relative uncertainty of $k$-th ($k=1,2,3,4$) order moment.
Our results are a significant step towards using a quantum sensing network with security inbuilt.
\end{abstract}
\maketitle
\section{Introduction}
Quantum properties such as superposition and entanglement are 
considered as resources for quantum information processing
\cite{shor1999polynomial,
PhysRevLett.79.325,
PhysRevLett.103.150502, 
vandersypen2001experimental, 
BENNETT20147,
bennett1992experimental,
RevModPhys.74.145}.
A quantum computer performs certain types of calculations much faster than classical computers
\cite{shor1999polynomial,
PhysRevLett.79.325,
PhysRevLett.103.150502, 
vandersypen2001experimental, 
BENNETT20147,
bennett1992experimental,
RevModPhys.74.145}.
Quantum cryptography provides a secure communication
\cite{BENNETT20147, 
bennett1992experimental,
RevModPhys.74.145}.
A concept to combine the quantum computation and quantum cryptography was proposed, which is called the blind quantum computation (BQC)
\cite{broadbent2009universal,
PhysRevA.87.050301,
PhysRevA.93.052307, 
barz2012demonstration,
greganti2016demonstration}.
In BQC, 
we assume that a client can perform only simple quantum operations such as single-qubit operations. 
The client delegates the computation to a server who possesses a quantum computer.
The BQC allows the client to
perform a quantum computation on the server side without revealing the input, the output, and the algorithm to the server.
Quantum anonymous communication allows users to exchange messages anonymously where the identity of the sender is hidden from each other.
There is a famous classical anonymous communication technology called
Tor (The Onion Routing)\cite{dingledine2004tor}.
However,
the Tor protocol relies on the public key encryption, 
and so it is vulnerable to attacks by quantum computers.
 On the other hand,
 Quantum anonymous communication is known to be secure even if it is attacked by quantum computers
\cite{christandl2005quantum,
unnikrishnan2019anonymity,
PRXQuantum.1.020325,
hahn2020anonymous,
khan2020quantum,
thalacker2021anonymous,
PRXQuantum.3.040306,
PhysRevA.101.062311,
PhysRevA.97.032345,
PhysRevA.98.052320,
gong2022anonymous,
newbrassard2007anonymous,
wang2010economical}.
Quantum sensing is one of the other applications \cite{
maze2008nanoscale,
balasubramanian2008nanoscale,
taylor2008high,
kominis2003subfemtotesla,
bal2012ultrasensitive,
eldredge2018optimal,
proctor2018multiparameter}.
When we create a superposition between a ground state and excited state of the qubit, 
an external magnetic field provides a relative phase between them.
By measuring the quantum state, 
we can estimate the amplitude of the field.
There have been many efforts in both theory and experiment regarding
quantum sensors
\cite{
maze2008nanoscale,
balasubramanian2008nanoscale,
taylor2008high,
kominis2003subfemtotesla,
bal2012ultrasensitive,
eldredge2018optimal,
proctor2018multiparameter}.
It is known that we can measure the magnetic field, electric field, and temperature by using quantum sensors
\cite{degen2017quantum, 
budker2007optical, 
balasubramanian2008nanoscale,
maze2008nanoscale, 
dolde2011electric, 
neumann2013high}.
By using quantum sensors, it is expected that we can detect small fields in a local region.
Such detection is useful in materials science, medical science, and biology
\cite{
mitchell2020colloquium,
schirhagl2014nitrogen,
barry2016optical}.
It is possible to improve the sensitivity of quantum sensors by using an entanglement
\cite{PhysRevA.46.R6797, 
PhysRevLett.79.3865,
PhysRevA.84.012103,
PhysRevLett.109.233601, 
jones2009magnetic}.
Also, 
there are numerous applications of a quantum sensor network where quantum sensors are located in distant places.
For example, 
we can efficiently estimate the magnetic-field distribution by using entanglement between them
\cite{
PhysRevX.11.031009,
brida2010experimental,
perez2012fundamental,
baumgratz2016quantum,
komar2014quantum}.
Importantly, 
we can use quantum sensing for magnetoencephalography
\cite{xia2006magnetoencephalography}.
This means that we could collect private information via quantum sensing in the future.
Therefore, 
it is important to realize a quantum sensing network with security
inbuilt. 
There were many attempts to hybridize quantum cryptography and quantum sensing
\cite{
degen2017quantum,
giovannetti2001quantum,
giovannetti2002quantum,
giovannetti2002positioning,
chiribella2005optimal,
chiribella2007secret,
huang2019cryptographic,
xie2018high,
PhysRevA.105.L010401,
giovannetti2006quantum,
giovannetti2011advances,
pirandola2020advances,
PhysRevA.104.062610,
yin2020experimental,
PhysRevX.11.031009,
brida2010experimental,
perez2012fundamental,
baumgratz2016quantum,
komar2014quantum,
maletinsky2012robust,
schirhagl2014nitrogen,
takeuchi2019quantum,
kasai2022anonymous}.
Recently, 
for the purpose to add security in a quantum sensing network, 
Shettell {\it et al.} proposed a method to estimate a linear combination of
magnetic fields at distant places without knowing the individual value of 
the magnetic fields
\cite{shettell2022private}.
They used Greenberger-Horne-Zeilinger (GHZ) states to obtain the estimation of
the linear combinations.
Also, 
they studied the relationship between fidelity and anonymity.
The definition of anonymity is that 
we cannot estimate the individual value of the magnetic field at each place.
However,
since only linear combinations 
of magnetic fields can be estimated in the previous approach, 
we cannot obtain the higher-order moments such as variance, skewness, and kurtosis.
Here, we propose a method to obtain the higher-order moments 
without knowing each value of the magnetic fields
by using the quantum sensing network.
Suppose that there are $L$ sensor holders at 
distant places to measure local magnetic fields.
We define the senders as the sensor holders.
Each sender has a qubit to interact with each magnetic field, 
and different senders are located at different places.
The senders share an entanglement between them, 
and let their quantum sensors interact with local magnetic fields.
By measuring specific observables, we can directly estimate the higher-order moments from the measurement results.
Moreover, by calculating quantum Fisher information just before the measurements, we show that we cannot obtain any information about the individual magnetic fields in the limit of a large $L$.
The structure of our paper is as follows. 
In section 2,
we review a quantum sensing protocol.
In section 3,
we introduce our method to measure the higher-order moments.
In section 4,
we explain how our method is secure in the sense that the individual values of
the magnetic fields cannot be estimated.
Finally, the paper is summarised and concluded in section 5 
and two appendices were included to add more technical details.
\section{Magnetic-field sensing with a qubit}
Let us review how to measure the amplitudes of the
magnetic fields with qubits \cite{degen2017quantum}.
\subsection{The dynamics of a qubit under the magnetic fields}
The dynamics of the qubit interacting with the magnetic fields are 
described as follows. 
Let us define $ \hat{\sigma}_{z}$ and $ \hat{\sigma}_{x}$ be the
standard Pauli operators. 
The Hamiltonian of the qubit is represented as
\begin{eqnarray}
\hat{H}
&=&
\frac{\omega}{2} \hat{\sigma}_{z}
.
\end{eqnarray}
We assume that the resonant frequency $\omega$ is
proportional to the applied magnetic fields. 
We choose $ \hat{\rho}_{0} = \ket{+}\bra{+} $,
an eigenstate of $\hat{\sigma}_{x}$, as the initial state. 
The state is evolved 
by the unitary operator $\hat{U} = e^{-i\hat{H} t}$. 
It is worth mentioning that $t$ denotes the interaction time with the
magnetic fields, and so $t$ is a known parameter. 
Throughout this paper, 
we assume $\hbar=1$.
After the time evolution, we obtain the quantum state as follows.
\begin{eqnarray}
{}
&{}&
\hat{\rho}_{0,\omega}
 \nonumber \\
&=&
\hat{U}
\hat{\rho}_{0}
\hat{U}^{\dagger}
\nonumber \\
&=&
\biggl\{
 \frac{1}{\sqrt{2}} \bigl( \ket{0} + e^{ i\omega t } \ket{1} \bigr)
\biggr\}
\biggl\{ 
 \frac{1}{\sqrt{2}} \bigl( \bra{0} + e^{ -i\omega t } \bra{1} \bigr)
\biggr\}
\nonumber \\
\end{eqnarray}
The relative phase is encoded in the quantum state, 
and the phase contains information on the resonant frequency
(corresponding to the applied magnetic fields).
\subsection{Evaluation of moments}
Let us define the $k$-th moment as
\begin{eqnarray}
\braket{ \omega^{k} }_{E}
\equiv
\frac{1}{L}
\sum _{l=0}^{L-1} (\omega_{l})^k
\end{eqnarray}
where $ \omega_{l}$ 
 denotes the resonant frequency at the site $l$.
Here, $L$ denotes the number of sites.
We define a characteristic function $B(t)$ as follows.
\begin{eqnarray}
B(t)
&\equiv&
\frac{1}{L}
\sum_{l=0}^{L-1}
e^{i \omega_{l} t}
\end{eqnarray}
The real and imaginary parts of the characteristic function $B(t)$ are represented as
\begin{eqnarray}
\begin{cases}
Re(B(t))
=
\sum_{k=0}^{\infty}
t^{2k} 
\frac
{ (-1)^{k} }{(2k)!}
\braket{ \omega^{2k} }_{E}
\\
Im(B(t))
=
\sum_{k=0}^{\infty}
t^{2k+1} 
\frac
{ (-1)^{k} }{(2k+1)!}
\braket{ \omega^{2k+1} }_{E}
.
\end{cases}
\end{eqnarray}
The moments of $\omega_l$ are represented as follows.
\begin{eqnarray}
\begin{cases}
\braket{ \omega^{2k} }_{E}
=
(-1)^{k} 
\frac
{ \partial^{2k} Re(B(t)) }
{ \partial t^{2k}}
\biggl|_{t=0}
\\
\braket{ \omega^{2k+1} }_{E}
=
(-1)^{k} 
\frac
{ \partial^{2k+1} Im(B(t)) }
{ \partial t^{2k+1}}
\biggl|_{t=0}
\end{cases}
\end{eqnarray}
To calculate the $2k$-th ($(2k+1)$-th) moment,
we need to differentiate the real (imaginary) part of the characteristic function with respect to the time $t$.
However,
it is not straightforward to obtain the value of the differentiation from an experiment.
So,
we consider finite differences of $Re(B(t))$ and $Im(B(t))$ at $t=0$. 
Especially, 
we adopt a forward difference as follows with using finite positive time $\Delta t$ .
\begin{eqnarray}
\begin{cases}
\frac
{ \Delta^{2k} Re(B(t)) }
{ \Delta t^{2k}}
\biggl|_{t=0}
\\
=
\frac{1}
{ ( \Delta t)^{2k} }
\sum_{l=0}^{2k}
(-1)^{l}
\binom{2k}{l}
Re(B( l \Delta t ))
\\
\frac
{ \Delta^{2k+1} Im(B(t)) }
{ \Delta t^{2k+1}}
\biggl|_{t=0}
\\
=
\frac{1}
{ ( \Delta t)^{2k+1} }
\sum_{l=1}^{2k+1}
(-1)^{l+1}
\binom{2k+1}{l}
Im(B( l \Delta t ))
\end{cases}
\end{eqnarray}
We define an estimated value of the moment from the finite difference as follows.
\begin{eqnarray}
\begin{cases}
\braket{ \omega^{2k} }_{E, \Delta}
\equiv
(-1)^{k} 
\frac
{ \Delta^{2k} Re(B(t)) }
{ \Delta t^{2k} }
\biggl|_{t=0}
\\
\braket{ \omega^{2k+1} }_{E, \Delta}
\equiv
(-1)^{k} 
\frac
{ \Delta^{2k+1} Im(B(t)) }
{ \Delta t^{2k+1} }
\biggl|_{t=0}
\end{cases}
\end{eqnarray}
 Considering the limit of small $\Delta t$,
 $\braket{ \omega^{2k} }_{E, \Delta}$
($\braket{ \omega^{2k+1} }_{E, \Delta}$)
 coincides with $\braket{ \omega^{2k} }_{E}$ ( $\braket{ \omega^{2k+1} }_{E}$).
\section{Anonymous estimation protocol of moments}
\subsection{Definitions and model}
\begin{figure*}[htbp]
\includegraphics[height=5.0cm,width=15cm]
{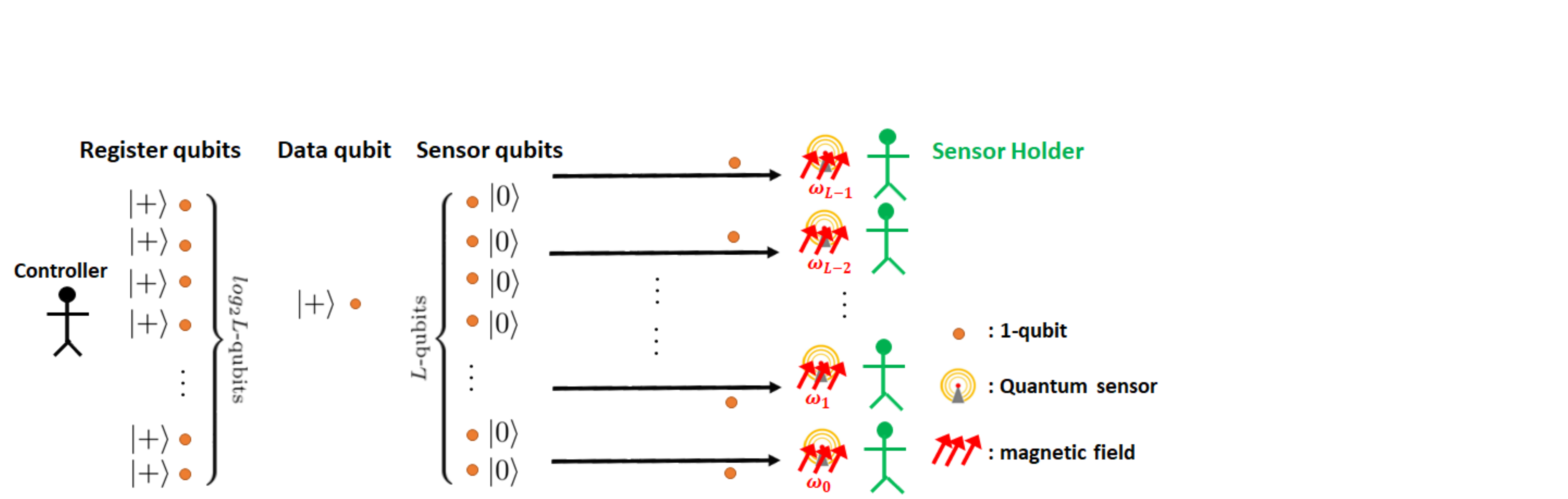}
\caption{
The schematic of our protocol
}
\label{figure:protocol-flow}
\end{figure*}
We explain our method to measure $\braket{ \omega^{2k} }_{E, \Delta}$
and
$\braket{ \omega^{2k+1} }_{E, \Delta}$.
For this purpose, we define the distributer, 
the controller and the sensor holder.
In Figure \ref{figure:protocol-flow},
we show a schematic of our method.
We assume that there is no decoherence. 
Also, 
we assume that we can perform ideal gate operations and ideal measurements.
\begin{enumerate}
\item The distributer \\

The role of the distributer is to distribute
the product states to the controller,
which will be defined later.
Prepare $M_{\rm{C}}=k+1$ copies of a state per a single cycle. This state is composed of register qubits, a data qubit, and sensor qubits. The number of register (sensor) qubits is ${\rm{log}}_{2} L$ ($L$). The register qubit is used to specify the site of the sensor holders. 
The sensor qubit interacts with magnetic fields, 
and acquires the relative phase due to the magnetic fields.
The data qubit stores the information of the relative phases acquired by the sensor qubits. 
We will show the mathematical definition later.
\item The controller \\

The controller performs a unitary operation 
on the qubits received from the distributer,
and then sends the state to the sensor holder which will be defined later.
After the sensor holders let the sensor qubits interact with magnetic fields, 
the controller receives the state from the sensor holders,
and performs a POVM measurement on the state. 
We assume that, 
since the controller is located far from the places to generate magnetic fields, 
any operation by the controller is not affected by magnetic fields.
\item The sensor holder \\

The place where each sensor holder is located generates unknown magnetic fields.
The sensor holder receives the state from the controller, 
lets the qubit interact with the magnetic fields at each place, 
and sends the state back to the controller.
In our protocol,
there are $L$ places that generate magnetic fields.
There is a sensor holder corresponding to each place.
Let the set of all sensor holders as 
$V = \{0,1,2,\ldots, L-1 \}$.
We define $w_{l}$ as the amplitude of the
magnetic field at the place 
where a sensor holder $l \in V$ is located. 
\end{enumerate}
\subsection{Protocol}
\label{sec:protocol_flow}
We define the CSWAP operation on the $j$-th copy as follows.
\begin{eqnarray}
\hat{U}_{\rm{CSWAP}}^{(j)}
&\equiv&
\sum_{l=0}^{L-1}
\ket{l}_{\rm{r},\textit{j}} \bra{l}
\otimes
\hat{U}_{\rm{SWAP}}^{(j)}(\rm{d}, \textit{l})
\end{eqnarray}
Here, 
$\hat{U}_{\rm{SWAP}}^{(j)}(\rm{d},\textit{l})$ denotes a SWAP operation 
between the data qubit and the $l$-th sensor qubit of the $j$-th copy.
The unitary evolution of the sensor qubit of 
$j$-th copy under the effect of magnetic field is described as follows.
\begin{eqnarray}
\label{eq:all_time_evolution}
\hat{U}_{B}^{(j)}(t)
\equiv
\bigotimes_{l=0}^{L-1}
e^{ -\frac{1}{2} 
i \omega_{l}t \hat{\sigma}_{z}^{(\rm{s},\textit{l},\textit{j})} }
\end{eqnarray} 
$\hat{\sigma}_{z}^{({\rm{s}},l,j)}$ denotes a Pauli matrix of the $l$-th sensor
qubit on the $j$-th copy.
We define the observable to be measured as follows.
\begin{eqnarray}
\hat{C} ( \Delta t, k)
&\equiv&
\frac{1}
{ ( \Delta t)^k }
\sum_{j=0}^{k}
(-1)^{k+j}
\binom{k}{j}
\hat{A}_{j}
\end{eqnarray}
$\hat{A}_{j}$ denotes an observable on the $j$-th copy, 
which will be defined later.
We provide an overview of our protocol.
\begin{oframed}
Init:
Specify $\Delta t$ and $k$. \\
Goal: 
The controller estimates the value of 
$\braket{ \omega^{k} }_{E}$.
\begin{enumerate}
\item Consider the following pure state of the $j$-th copy. \\
\begin{eqnarray}
\ket{\phi_0}_{j}
&\equiv&
\ket{+^{ \log_{2} L}}_{\rm{r},\textit{j}} 
\ket{+}_{\rm{d},\textit{j}} \ket{0^{ L}}_{\rm{s},\textit{j}}
\nonumber \\
&=&
\frac{1}{\sqrt{L} }
\sum_{l=0}^{L-1}
\ket{l}_{\rm{r},\textit{j}} \ket{+}_{\rm{d},\textit{j}} 
\ket{0^{ L}}_{\rm{s},\textit{j}}
\end{eqnarray}
where $\ket{l}_{\rm{r},\textit{j}}$,
$ \ket{+}_{\rm{d},\textit{j}} $, 
and $\ket{0^{ L}}_{\rm{s},\textit{j}}$
denote the state of the register qubits, data qubit, 
sensor qubits, respectively.
Also, let us define an initial state as
$
\hat{\rho}_{init,j}
\equiv
\ket{\phi_0}_{j} \bra{\phi_0}
$ for $j=0,1,\ldots,k$.
Then,
the distributer sends $(k+1)$ copies of
the initial state $\hat{\rho}_{init,j}$ to the controller. 
The controller has the following state.
\begin{eqnarray}
{}
&{}&
\hat{ \rho}_{init,\rm{ALL}}
\nonumber \\
&\equiv&
\hat{\rho}_{init,0} 
\otimes 
\hat{\rho}_{init,1}
\otimes 
\nonumber \\
&\cdots&
\otimes 
\hat{\rho}_{init,k-1}
\otimes 
\hat{\rho}_{init,k}
\end{eqnarray}
Here, the $j$-th subsystem corresponds to the $j$-th copy.
\item The controller performs the following unitary operation.\\
\begin{eqnarray}
{}
&{}&
\hat{U}_{\rm{CSWAP},all}
\nonumber \\
&\equiv&
\hat{U}_{\rm{CSWAP}}^{(0)} 
\otimes 
\hat{U}_{\rm{CSWAP}}^{(1)}
\otimes 
\nonumber \\
&\cdots&
\otimes 
\hat{U}_{\rm{CSWAP}}^{(k-1)}
\otimes 
\hat{U}_{\rm{CSWAP}}^{(k)}
\end{eqnarray}
\item For all copies,
the controller sends the $l$-th sensor qubit to the $l$-th sensor holder.\\
\item Each sensor holder lets their sensor qubits interact with the magnetic fields. 
The interaction time at the $j$-th copy is $j\Delta t$. 
The unitary operator induced by the magnetic fields is described as follows.
\begin{eqnarray}
{}
&{}&
\hat{U}_{B,all}
\nonumber \\
&\equiv&
\hat{U}_{B}^{(0)}(0)
\otimes 
\hat{U}_{B}^{(1)}( \Delta t)
\otimes 
\nonumber \\
&\cdots&
\otimes 
\hat{U}_{B}^{(k-1)}( (k-1) \Delta t)
\otimes 
\hat{U}_{B}^{(k)}( k\Delta t)
\nonumber \\
\end{eqnarray}
\item All sensor holders send the sensor qubits back to the controller.\\

\item The controller performs $\hat{U}_{\rm{CSWAP},all}$.\\

\item The controller measures the observable $\hat{C} ( \Delta t, k)$.\\

\item Repeat the above steps $N$ times.\\
\end{enumerate}
\end{oframed}
\section{Analytics of estimation protocol}
\subsection{Properties}
The purpose of our protocol is to estimate the value of $\braket{ \omega^{k} }_{E}$ without knowing the value of $\omega_{l}$. 
If an eavesdropper (Eve) performed some POVM measurements 
on the state at the end of step 4, 
Eve could obtain the information of $\omega_{l}$. 
We consider this attack.
So we consider our protocol as anonymity when Eve cannot obtain any information about $\omega_{l}$ by performing any POVM measurement at the end of the step 4.
Such anonymity is important for the following reason. After the step 6, 
the controller still has the quantum state.
Since this quantum state contains the information of $\omega_{l}$, 
the controller should keep the quantum state in a quantum memory forever to prevent information leakage, 
which is practically difficult. 
It is worth mentioning that, in such a case, 
the controller cannot naively initialize the quantum state by measurements
because such an initialization could lead to information leakage.
On the other hand, if our protocol satisfies anonymity, 
any POVM measurement for the initialization does not leak any information of $\omega_{l}$, 
and so the quantum memory to keep the quantum state is not required.
\subsubsection{Correctness}
\label{sec:correctness}
We evaluate estimation uncertainty of $\braket{ \omega^{k} }_{E}$ 
on our method.
Let us consider a state
$
\hat{ \rho}_{all,fin}
(\Delta t, k)
$
just before measuring the observable $\hat{C} ( \Delta t, k)$ at the step 7.
For $k=2m$,
the observable $\hat{A}_j$, which acts on the $j$-th copy, is given as 
$
\hat{A}_{j} 
\equiv 
\ket{+^{log_2 L}}_{\rm{r},\rm{j}} \bra{+^{ log_2 L}}
\otimes 
\hat{\sigma}_{x}^{(\rm{d},\rm{j})}
$. 
The expectation value of $\hat{A}_{j}$ is given as follows.
\begin{eqnarray}
{}
&{}&
A(t)
=
\braket{
\hat{A}_{j}
}_{E}
\nonumber \\
&=&
\braket{
\ket{+^{ log_2 L}}_{\rm{r},\textit{j}}
\bra{+^{ log_2 L}} 
\otimes
\hat{\sigma}_{x}^{(\rm{d},\textit{j})}
}_{E}
\nonumber \\
&=&
\mathrm{Tr} 
\{ 
\hat{ \rho}_{fin.j}(t)
(
\ket{+^{ log_2 L}}_{\rm{r},\textit{j}}
\bra{+^{ log_2 L}} 
\otimes
\hat{\sigma}_{x}^{(\rm{d},\textit{j})}
)
\}
\nonumber \\
&=&
\mathrm{Tr} 
\{ 
\ket{\phi_4(t)}_{j} \bra{\phi_4(t)}
(
\ket{+^{ log_2 L}}_{\rm{r},\textit{j}}
\bra{+^{ log_2 L}} 
\otimes
\hat{\sigma}_{x}^{(\rm{d},\textit{j})}
)
\}
\nonumber \\
&=&
\braket{ \phi_4(t) |
(
\ket{+^{ log_2 L}}_{\rm{r},\textit{j}}
\bra{+^{ log_2 L}} 
\otimes
\hat{\sigma}_{x}^{(\rm{d},\textit{j})}
)
| \phi_4(t) }
\nonumber \\
&=&
\bra{ \phi_4(t)} \cdot 
\ket{+^{ log_2 L}}_{\rm{r},\textit{j}}
\hat{\sigma}_{x}^{(\rm{d},\textit{j})}
\bra{+^{ log_2 L}}_{\rm{r},\textit{j}}
\cdot \ket{ \phi_4(t)} 
\nonumber \\
&=&
\bra{A(t)} _{\rm{d}}
\cdot
\hat{\sigma}_{x}^{(\rm{d},\textit{j})} 
\cdot
\ket{A(t)}_{\rm{d}}
=
Re(B(t)) 
\end{eqnarray}
where the definition of $\ket{ \phi_4(t)}$ ( $|A(t)\rangle $ ) is
Eq. \eqref{eq:phi_4_ket} (Eq. \eqref{eq:Aket}) .
For $k=2m$,
the observable $\hat{C} ( \Delta t, k)$ is described as follows.
\begin{eqnarray}
{}
&{}&
\hat{C}(\Delta t,k=2m) 
 \nonumber \\
&\equiv&
\frac{1}
{ ( \Delta t)^{2m} }
\sum_{j=0}^{2m}
(-1)^{j}
\binom{2m}{j}
(
\ket{+^{ log_{2} L}}_{\rm{r},\textit{j}}
\bra{+^{ log_{2} L}} 
 \nonumber \\
&\otimes&
\hat{\sigma}_{x}^{(\rm{d},\textit{j})}
)
\end{eqnarray}
We calculate the expectation value and the uncertainty of $ \hat{C}(\Delta t,k=2m) $ as follows.
Let us define the expectation value and variance as
$C(t=0,\Delta t, k=2m)
\equiv
\braket{ \hat{C}( \Delta t, k=2m) }_{E}$ and
$\delta (C^{2}) (t=0,\Delta t, k=2m) \equiv \braket{ \hat{C}^2( \Delta t, k=2m) }_{E}-\braket{ \hat{C}( \Delta t, k=2m) }_{E}^2$
, respectively. 
We obtain the following.
\begin{eqnarray}
{}
&{}&
C(t=0,\Delta t, k=2m)
\nonumber \\
&=&
\mathrm{Tr} 
(
\hat{ \rho}_{all,fin}(\Delta t,k=2m) 
\hat{C}(\Delta t,k=2m) 
)
\nonumber \\
&=&
\frac
{ \Delta^{2m} A(t)}
{ \Delta t^{2m} }
\biggl|_{t=0}
=
\frac
{ \Delta^{2m} Re(B(t)) }
{ \Delta t^{2m} }
\biggl|_{t=0}
\nonumber \\
&=&
(-1)^{m} 
\braket{ \omega^{2m} }_{E, \Delta}
\end{eqnarray}
\begin{eqnarray}
{}
&{}&
\delta (C^{2}) (t=0,\Delta t, k=2m)
\nonumber \\
&=&
( C^{2} ) (t=0,\Delta t, k=2m)
\nonumber \\
&-&
\{C(t=0,\Delta t, k=2m)\}^{2}
\nonumber \\
&=&
\frac{1}{ 2( \Delta t)^{4m} }
\sum_{j=1}^{2m}
\binom{2m}{j}^{2}
[ 
1
\nonumber \\
&+&
\{ Im(B( j \Delta t )) \}^{2} 
-
\{ Re(B( j \Delta t )) \}^{2} 
]
\nonumber \\
\end{eqnarray}
As we decrease $\Delta t$, $\delta (C^{2}) (t=0,\Delta t, k=2m)$ increases.
Also,
we obtain the uncertainty of the estimation as follows.
\begin{eqnarray}
{}
&{}&
\braket{ 
(
\braket{ \omega^{2m} }_{E,\Delta ,est}
-
\braket{ \omega^{2m} }_{E}
)^2
}_{E}
 \nonumber \\
&=&
\biggl<
\biggl(
\frac
{ \Delta^{2m} A^{est}(t)}
{ \Delta t^{2m} }
\biggl|_{t=0}
-
\frac
{ d^{2m} A(t)}
{ d t^{2m} }
\biggl|_{t=0}
\biggr)^2
\biggr>_{E}
\nonumber \\
&=&
\frac{
\delta (C^{2}) (t=0,\Delta t, k=2m)
}{N}
\nonumber \\
&+&
\epsilon
(t=0, \Delta t, k=2m)^2
\nonumber \\
&=&
\frac{
\delta (C^{2}) (t=0,\Delta t, k=2m)
}{N}
\nonumber \\
&+&
( \braket{ \omega^{2m} }_{E, \Delta} 
-
\braket{ \omega^{2m} }_{E} )^2
\label{uncertaintyym}
\end{eqnarray}
where
$\braket{ 
(
\braket{ \omega^{2m} }_{E,\Delta ,est}
-
\braket{ \omega^{2m} }_{E}
)^2
}_{E}$
denotes the statistical average of 
$(\braket{ \omega^{2m} }_{E,\Delta ,est}
-
\braket{ \omega^{2m} }_{E}
)^2$
and 
$
\epsilon
(t=0, \Delta t, k=2m)
\equiv
\frac
{ \Delta^{2m} A(t)}
{ \Delta t^{2m} }
\bigl|_{t=0}
-
\frac
{ d^{2m} A(t)}
{ d t^{2m} }
\bigl|_{t=0}
=
(-1)^{m}
(
\braket{ \omega^{2m} }_{E, \Delta} 
-
\braket{ \omega^{2m} }_{E} 
)
$ denotes a systematic error.
It is worth mentioning that, due to the central limit theorem, 
the first term in the right-hand side of 
Eq. \eqref{uncertaintyym} decreases by $1/N$ as we increase $N$.
On the other hand, 
the second term in the right-hand side of Eq. \eqref{uncertaintyym} is independent of $N$. \par
When the order of the moment is odd,
we prepare $M_{\rm{C}}=2m+2$ $(m=0,1,\ldots)$ copies of state.
Then, we define
$
\hat{A}_{j} 
\equiv 
\ket{+^{log_2 L}}_{\rm{r},\rm{j}} \bra{+^{ log_2 L}}
\otimes 
\hat{\sigma}_{y}^{(\rm{d},\rm{j})}
$.
For $k=2m+1$, 
the expectation value of $\hat{A}_{j}$ is calculated as follows.
\begin{eqnarray}
{}
&{}&
A(t)
=
\braket{
\hat{A}_{j}
}_{E}
\nonumber \\
&=&
\braket{
\ket{+^{ log_2 L}}_{\rm{r},\textit{j}}
\bra{+^{ log_2 L}} 
\otimes
\hat{\sigma}_{y}^{(\rm{d},\textit{j})}
}_{E}
\nonumber \\
&=&
\mathrm{Tr} 
\{ 
\hat{ \rho}_{fin.j}(t)
(
\ket{+^{ log_2 L}}_{\rm{r},\textit{j}}
\bra{+^{ log_2 L}} 
\otimes
\hat{\sigma}_{y}^{(\rm{d},\textit{j})}
)
\}
\nonumber \\
&=&
\mathrm{Tr} 
\{ 
\ket{\phi_4(t)}_{j} \bra{\phi_4(t)}
(
\ket{+^{ log_2 L}}_{\rm{r},\textit{j}}
\bra{+^{ log_2 L}} 
\otimes
\hat{\sigma}_{y}^{(\rm{d},\textit{j})}
)
\}
\nonumber \\
&=&
\braket{ \phi_4(t) |
(
\ket{+^{ log_2 L}}_{\rm{r},\textit{j}}
\bra{+^{ log_2 L}} 
\otimes
\hat{\sigma}_{y}^{(\rm{d},\textit{j})}
)
| \phi_4(t) }
\nonumber \\
&=&
\bra{ \phi_4(t)} \cdot 
\ket{+^{ log_2 L}}_{\rm{r},\textit{j}}
\hat{\sigma}_{y}^{(\rm{d},\textit{j})}
\bra{+^{ log_2 L}}_{\rm{r},\textit{j}}
\cdot \ket{ \phi_4(t)} 
\nonumber \\
&=&
\bra{A(t)} _{\rm{d}}
\cdot
\hat{\sigma}_{y}^{(\rm{d},\textit{j})} 
\cdot
\ket{A(t)}_{\rm{d}}
=
Im(B(t)) 
\end{eqnarray}
The observable to be measured is given as follows.
\begin{eqnarray}
{}
&{}&
\hat{C} (\Delta t, k=2m+1)
 \nonumber \\
&\equiv&
\frac{1}
{ ( \Delta t)^{2m+1} }
\sum_{j=0}^{2m+1}
(-1)^{j+1}
\binom{2m+1}{j}
 \nonumber \\
&\cdot&
(
\ket{+^{ log_2 L}}_{\rm{r},\textit{j}}
\bra{+^{ log_2 L}} 
\otimes
\hat{\sigma}_{y}^{(\rm{d},\textit{j})}
)
\end{eqnarray}
Let us define the expectation value and variance as
$C(t=0,\Delta t, k=2m+1)
\equiv
\braket{ \hat{C}( \Delta t, k=2m+1) }_{E}$ and
$\delta (C^{2}) (t=0,\Delta t, k=2m+1) \equiv \braket{ \hat{C}^2( \Delta t, k=2m+1) }_{E}-\braket{ \hat{C}( \Delta t, k=2m+1) }_{E}^2$.
We obtain the following.
\begin{eqnarray}
{}
&{}&
C(t=0,\Delta t, k=2m+1)
 \nonumber \\
&\equiv&
\braket{ \hat{C} (\Delta t, k=2m+1) }_{E}
 \nonumber \\
&=&
\mathrm{Tr} 
(
\hat{ \rho}_{all,fin}(\Delta t, k=2m+1)
\hat{C} (\Delta t, k=2m+1) 
)
\nonumber \\
&=&
\frac
{ \Delta^{2m+1} A(t)}
{ \Delta t^{2m+1} }
\biggl|_{t=0}
\nonumber \\
&=&
\frac
{ \Delta^{2m+1} Im(B(t)) }
{ \Delta t^{2m+1} }
\biggl|_{t=0}
=
(-1)^{m} 
\braket{ \omega^{2m+1} }_{E,\Delta}
\end{eqnarray}
\begin{eqnarray}
{}
&{}&
\delta (C^{2}) (t=0,\Delta t, k=2m+1)
\nonumber \\
&=&
( C^{2} ) (t=0,\Delta t, k=2m+1)
\nonumber \\
&-&
\{ C(t=0,\Delta t, k=2m+1) \}^{2}
\nonumber \\
&=&
\frac{1}{ 2( \Delta t)^{4m+2} }
\biggl[
\binom{4m+2}{2m+1} + 1
\nonumber \\
&+&
\sum_{j=1}^{2m+1}
 \binom{2m+1}{j}^{2}
\bigl[
\{ Re(B( j \Delta t )) \}^{2} 
\nonumber \\
&-&
\{ Im(B( j \Delta t )) \}^{2} 
\bigr]
\biggr]
\end{eqnarray}
As we decrease $\Delta t$, $\delta (C^{2}) (t=0,\Delta t, k=2m+1)$ increases.
The uncertainty is calculated as follows.
\begin{eqnarray}
{}
&{}&
\braket{ 
(
\braket{ \omega^{2m+1} }_{E,\Delta ,est}
-
\braket{ \omega^{2m+1} }_{E}
)^2
}_{E}
\nonumber \\
&=&
\biggl<
\biggl(
\frac
{ \Delta^{2m+1} A^{est}(t)}
{ \Delta t^{2m+1} }
\biggl|_{t=0}
-
\frac
{ d^{2m+1} A(t)}
{ d t^{2m+1} }
\biggl|_{t=0}
\biggr)^2
\biggr>_{E}
\nonumber \\
&=&
\frac{
\delta (C^{2}) (t=0,\Delta t, k=2m+1)
}{N}
\nonumber \\
&+&
\epsilon
(t=0, \Delta t, k=2m+1)^2
\nonumber \\
&=&
\frac{
\delta (C^{2}) (t=0,\Delta t, k=2m+1)
}{N}
\nonumber \\
&+&
( \braket{ \omega^{2m+1} }_{E, \Delta} 
-
\braket{ \omega^{2m+1} }_{E} )^2
\end{eqnarray}
Also, we calculate the relative uncertainty for $k=2m$
as follows.
\begin{widetext}
\begin{eqnarray}
\label{eq:relative_uncertainty_at_even_order_moment}
{}
&{}&
\omega_{\text{ relative uncertainty}} 
(\Delta t, k=2m ) 
\nonumber \\
&\equiv&
\frac
{
\sqrt{
\braket{ 
(
\braket{ \omega^{2m} }_{E,\Delta ,est}
-
\braket{ \omega^{2m} }_{E}
)^2
}_{E}
}
}
{
\braket{ \omega^{2m} }_{E}
}
\nonumber \\
&=&
\frac
{1} { \braket{ \omega^{2m} }_{E} }
\sqrt{
\frac{
\delta (C^{2}) (t=0, \Delta t, k=2m )
}{N}
+
( \braket{ \omega^{2m} }_{E, \Delta} 
-
\braket{ \omega^{2m} }_{E} )^2
}
\nonumber \\
&=&
\frac
{1} { \braket{ \omega^{2m} }_{E} }
\sqrt{
\frac{
\delta (C^{2}) (t=0, \Delta t, k=2m )
}{N}
+
\biggl(
(-1)^{m}
\frac
{ \Delta^{2m} Re(B(t)) }
{ \Delta t^{2m}}
\biggl|_{t=0}
- 
\braket{ \omega^{2m} }_{E} 
\biggr)^2
}
\nonumber \\
&=&
\frac
{1} { \braket{ \omega^{2m} }_{E} }
\biggl[
\frac{1}{N}
\frac{1}{ 2( \Delta t)^{4m} }
\sum_{j=1}^{2m}
 \binom{2m}{j}^{2}
[ 
1
+
\{ Im(B( j \Delta t )) \}^{2} 
-
\{ Re(B( j \Delta t )) \}^{2} 
]
\nonumber \\
&+&
\biggl\{
\frac
{ (-1)^{m} }
{ ( \Delta t)^{2m} }
\sum_{j=0}^{2m}
(-1)^{j}
\binom{2m}{j}
Re(B( j \Delta t ))
- 
\braket{ \omega^{2m} }_{E} 
\biggr\}^2
\biggr]^{ \frac{1}{2} }
\end{eqnarray}
\end{widetext}
We calculate the relative uncertainty for $k=2m+1$
as follows.
\begin{widetext}
\begin{eqnarray}
\label{eq:relative_uncertainty_at_odd_order_moment}
{}
&{}&
\omega_{\text{ relative uncertainty}} 
( \Delta t, k=2m+1 ) 
\nonumber \\
&\equiv&
\frac
{
\sqrt{
\braket{ 
(
\braket{ \omega^{2m+1} }_{E,\Delta ,est}
-
\braket{ \omega^{2m+1} }_{E}
)^2
}_{E}
}
}
{
\braket{ \omega^{2m+1} }_{E}
}
\nonumber \\
&=&
\frac
{1} { \braket{ \omega^{2m+1} }_{E} }
\sqrt{
\frac{
\delta (C^{2}) (t=0, \Delta t, k=2m+1 )
}{N}
+
( \braket{ \omega^{2m+1} }_{E, \Delta} 
-
\braket{ \omega^{2m+1} }_{E} )^2
}
\nonumber \\
&=&
\frac
{1} { \braket{ \omega^{2m+1} }_{E} }
\sqrt{
\frac{
\delta (C^{2}) (t=0, \Delta t, k=2m+1 )
}{N}
+
\biggl(
(-1)^{m}
\frac
{ \Delta^{2m+1} Im(B(t)) }
{ \Delta t^{2m+1}}
\biggl|_{t=0}
- 
\braket{ \omega^{2m+1} }_{E} 
\biggr)^2
}
\nonumber \\
&=&
\frac
{1} { \braket{ \omega^{2m+1} }_{E} }
\biggl[
\frac{1}{N}
\frac{1}{ 2( \Delta t)^{4m+2} }
\sum_{j=0}^{2m+1}
\binom{2m+1}{j}^{2}
[ 
1
+
\{ Re(B( j \Delta t )) \}^{2} 
-
\{ Im(B( j \Delta t )) \}^{2} 
]
\nonumber \\
&+&
\biggl\{
\frac
{ (-1)^{m} }
{ ( \Delta t)^{2m+1} }
\sum_{j=1}^{2m+1}
(-1)^{j+1}
\binom{2m+1}{j}
Im(B( j \Delta t ))
- 
\braket{ \omega^{2m+1} }_{E} 
\biggr\}^2
\biggr]^{ \frac{1}{2} }
\end{eqnarray}
\end{widetext}
\begin{figure*}[t!]
\begin{tabular}{cc}
\begin{minipage}[t]{0.5\hsize}
\centering
\includegraphics[clip,width=0.8\textwidth]
{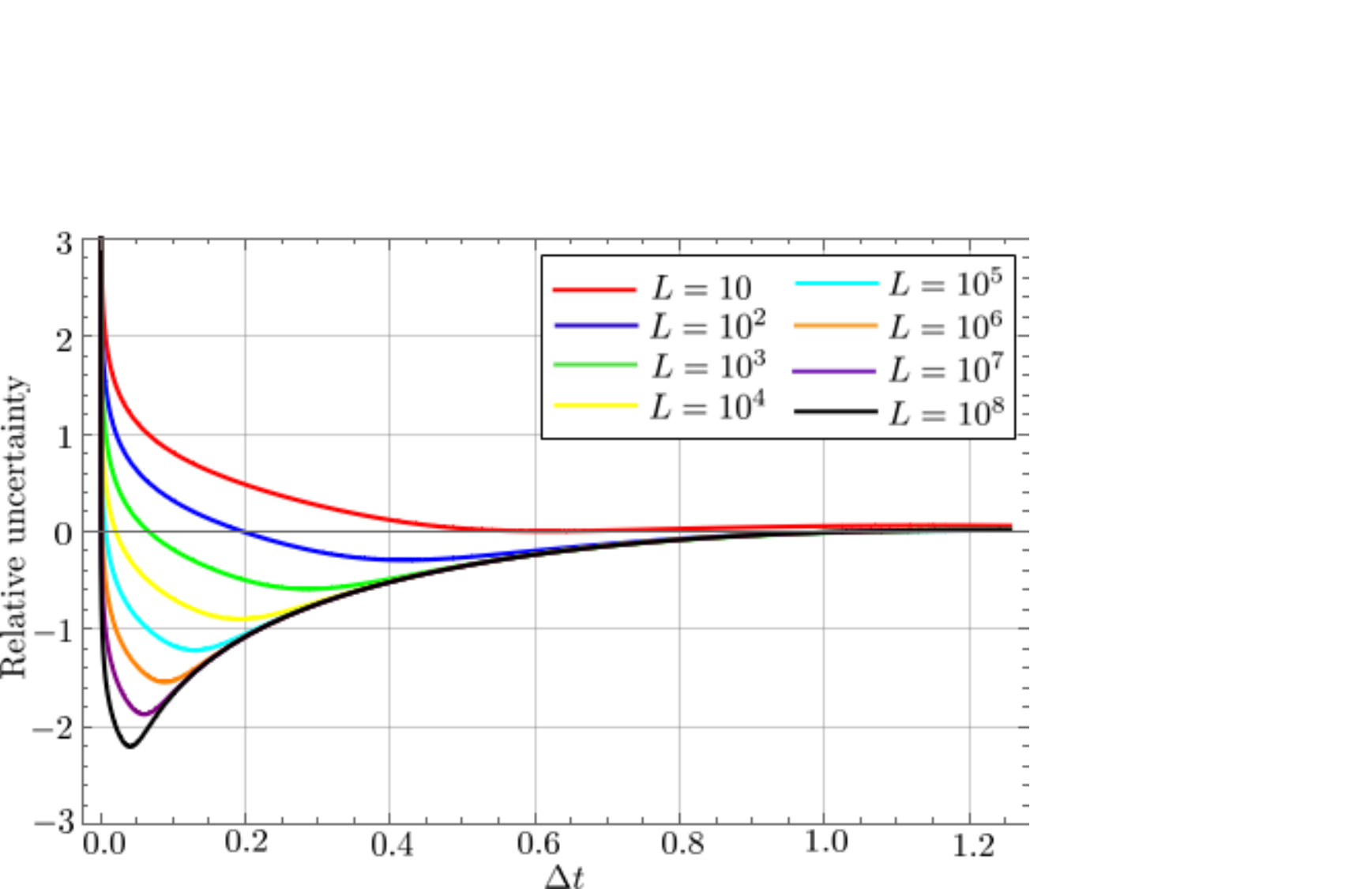}
\subfigure{
\label{fig:relative-uncertainty-1st-moment}
(a) $1$st moment }
\end{minipage} &
\begin{minipage}[t]{0.5\hsize}
\centering
\includegraphics[clip,width=0.8\textwidth]
{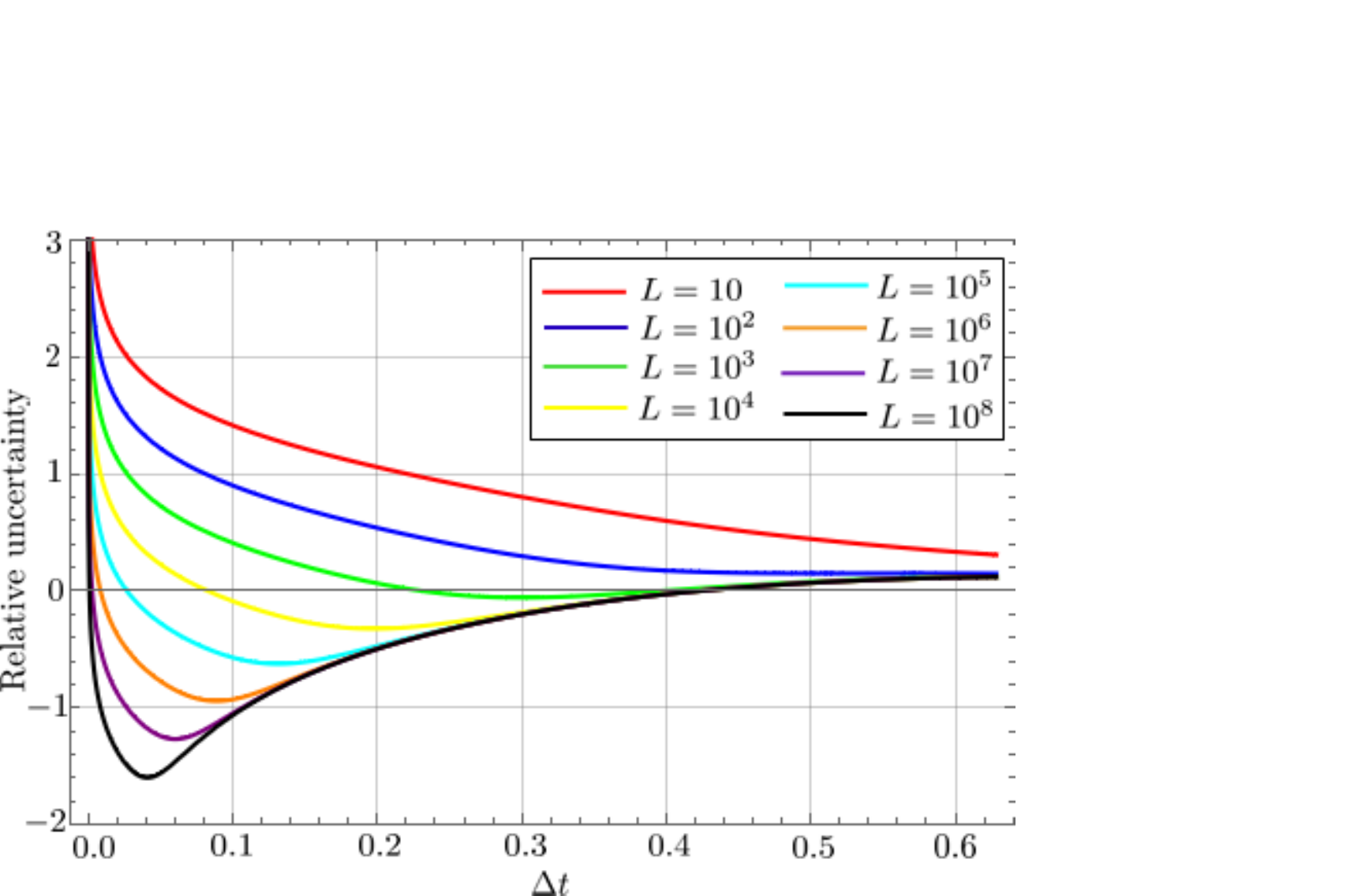}
\subfigure{(b) $2$nd moment}
\label{fig:relative-uncertainty-2nd-moment}
\end{minipage} \\
\begin{minipage}[t]{0.5\hsize}
\centering
\includegraphics[clip,width=0.8\textwidth]
{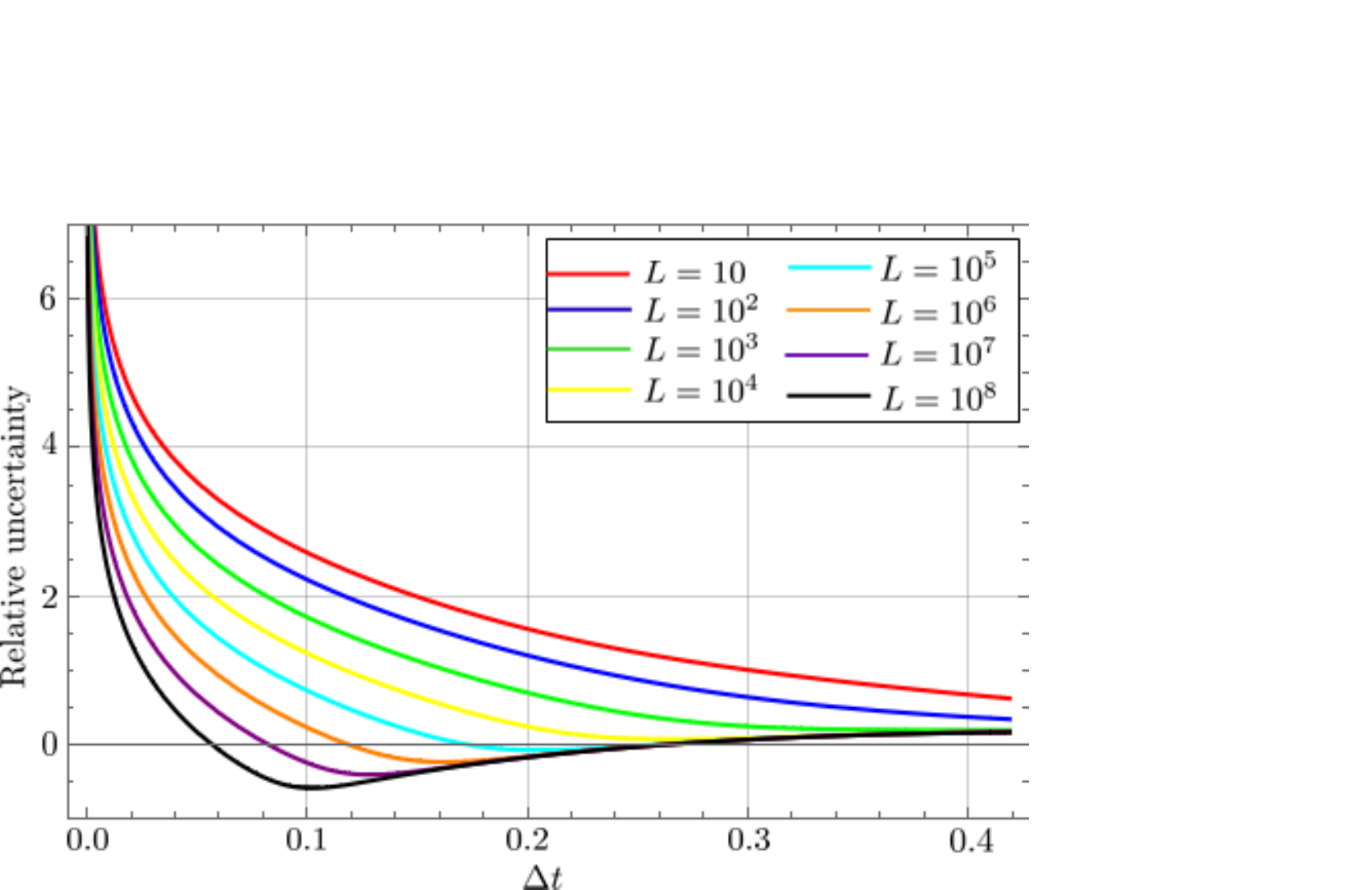}
\subfigure{(c) $3$rd moment }
\label{fig:relative-uncertainty-3rd-moment}
\end{minipage} &
\begin{minipage}[t]{0.5\hsize}
\centering
\includegraphics[clip,width=0.8\textwidth]
{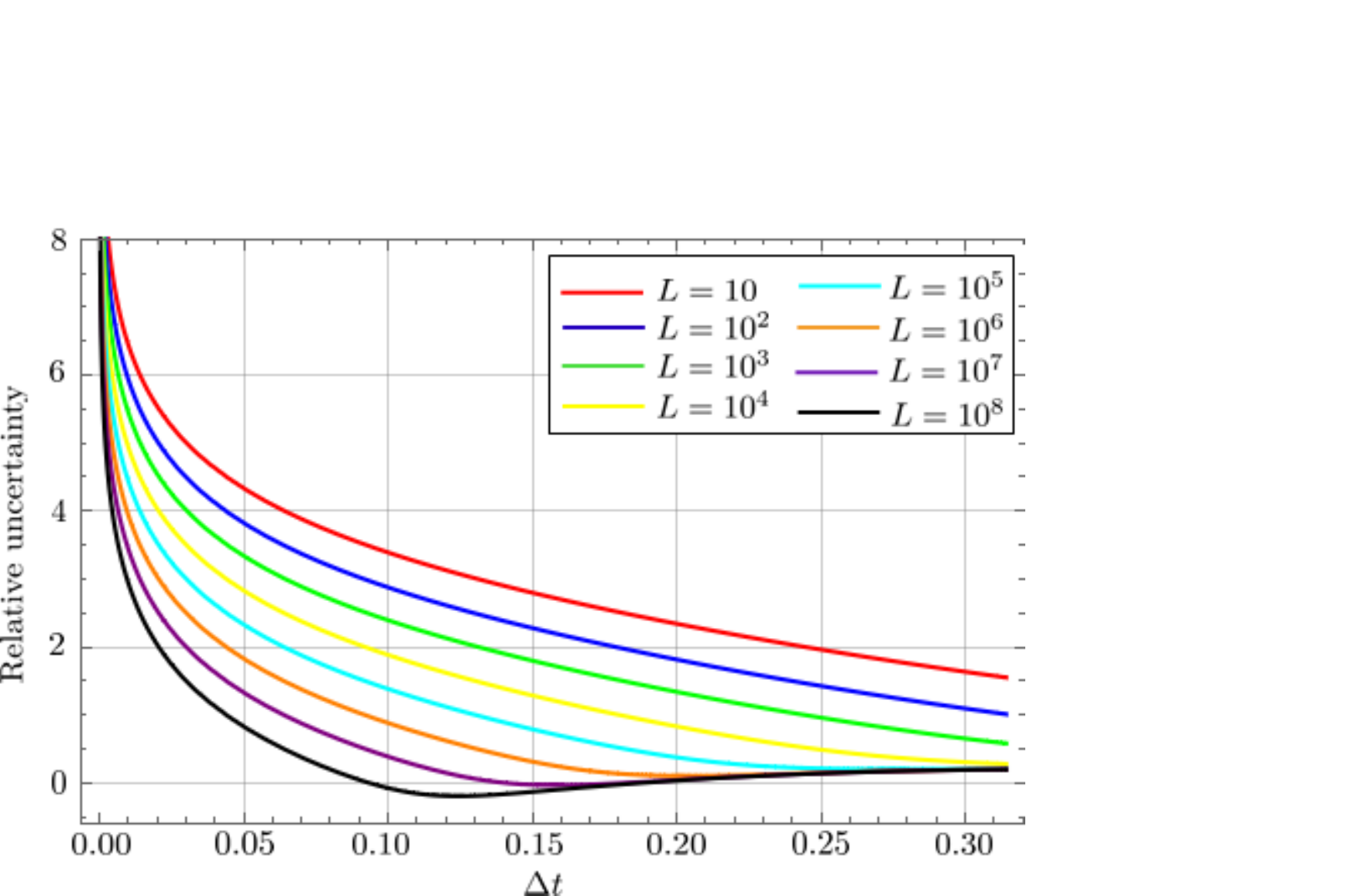}
\subfigure{(d) $4$th moment }
\label{fig:relative-uncertainty-4th-moment}
\end{minipage} \\
\begin{minipage}[t]{0.5\hsize}
\centering
\includegraphics[clip,width=0.8\textwidth]
{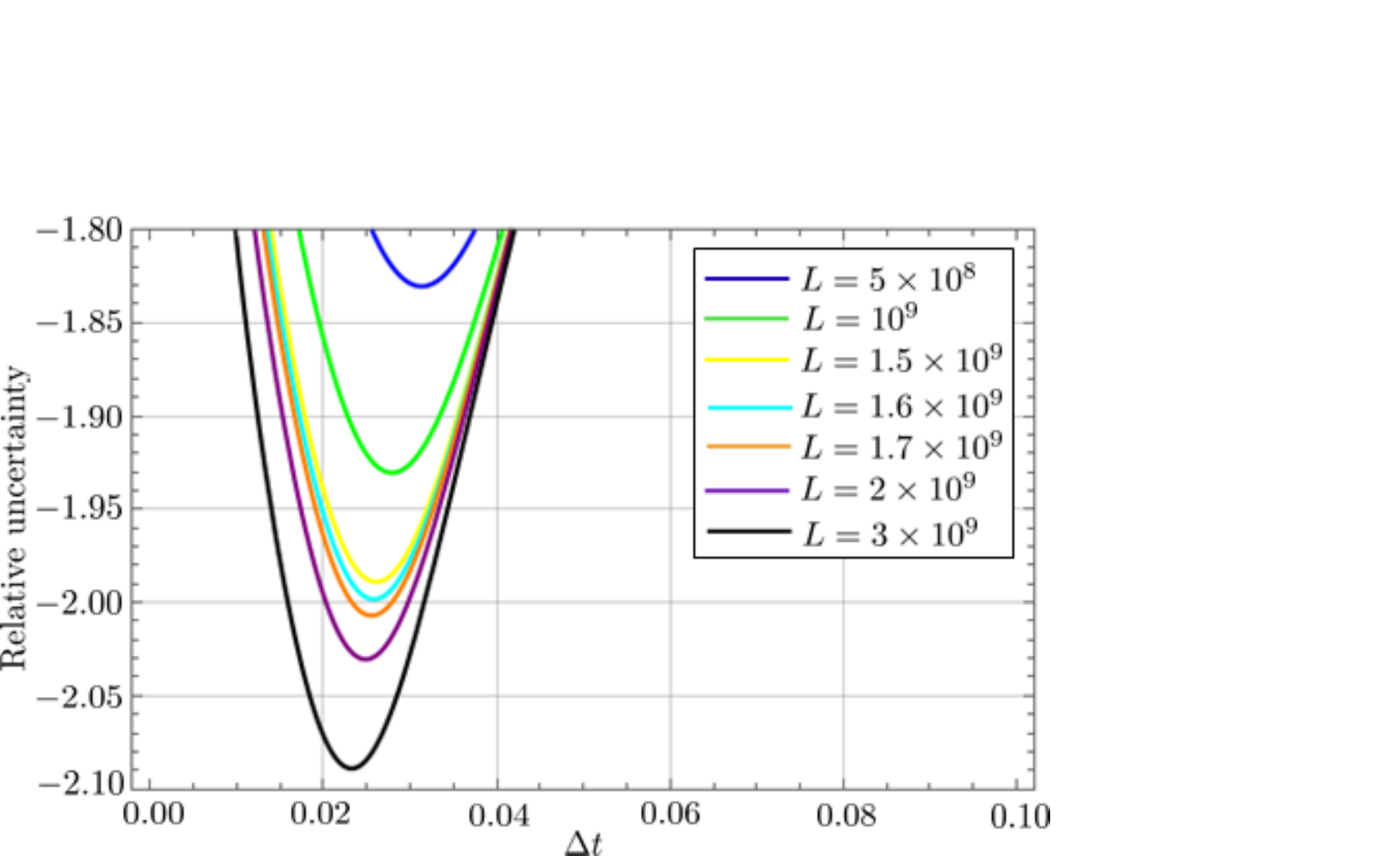}
\subfigure{(e) $2$nd moment }
\label{fig:relative-uncertainty-minus-2-at-2nd-moment}
\end{minipage} &
\end{tabular}
 \caption{
The plot of the relative uncertainty of the moment on log scale.
The horizontal axis is the time difference $\Delta t$ and
the vertical axis is relative uncertainty in log scale. 
}
 \label{fig:relative-uncertainty-of-moment}
 \end{figure*}
\begin{figure*}[t!]
\begin{tabular}{cc}
\begin{minipage}[t]{0.5\hsize}
\centering
\includegraphics[clip,width=0.8\textwidth]
{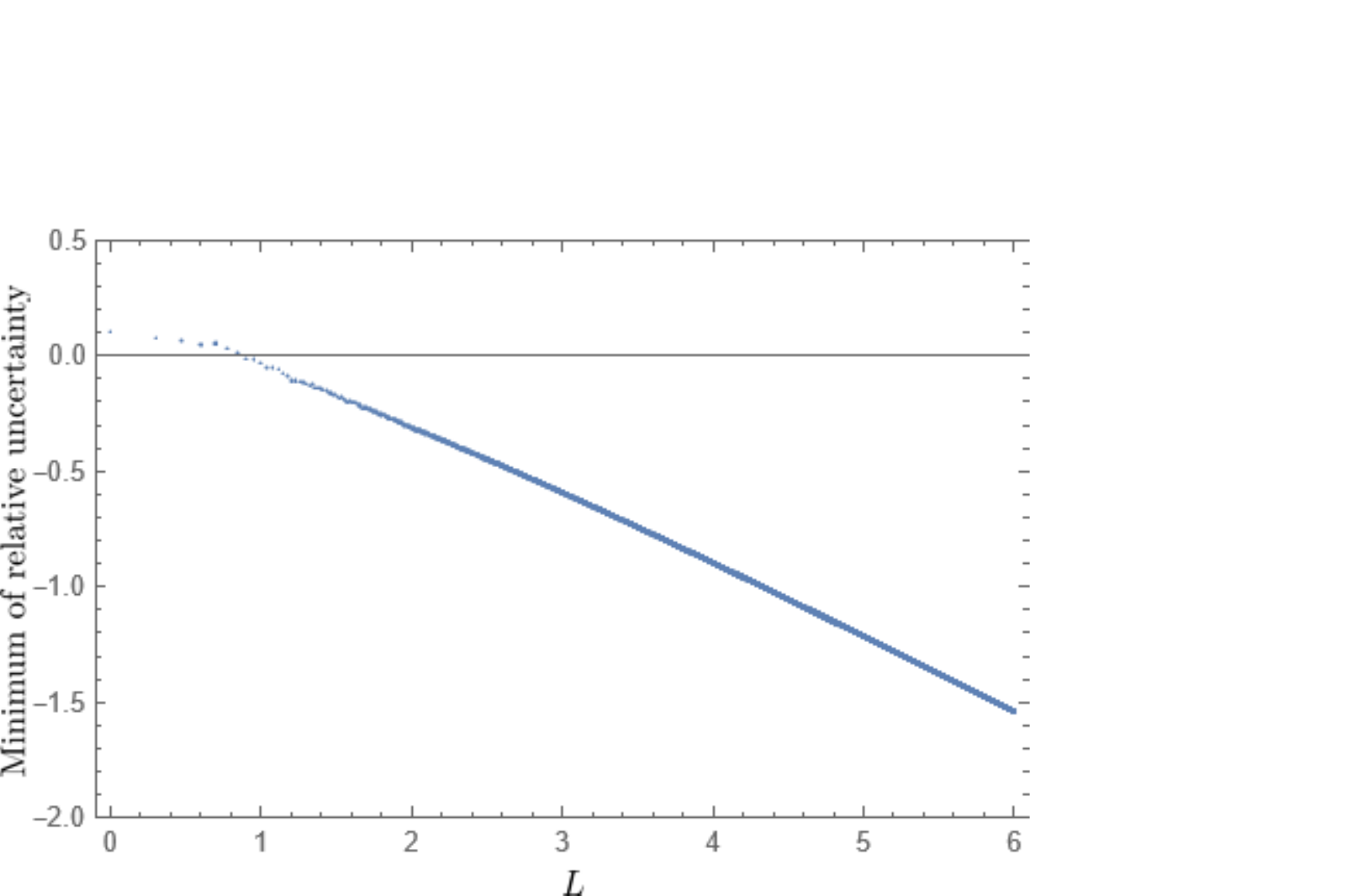}
\subfigure{(a) $1$st moment }
\label{fig:min-relative-uncertainty-1st-moment}
\end{minipage} &
\begin{minipage}[t]{0.5\hsize}
\centering
\includegraphics[clip,width=0.8\textwidth]
{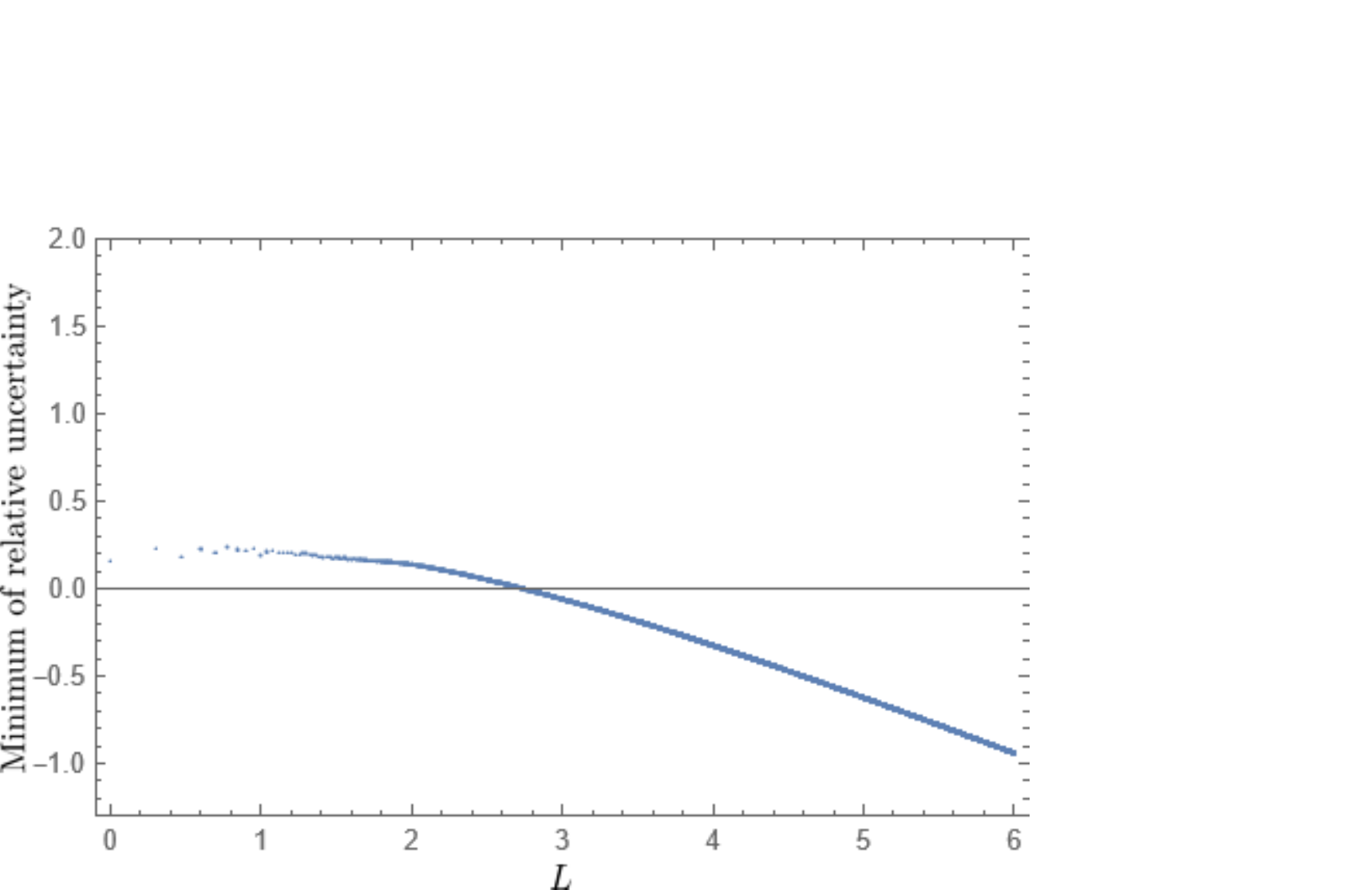}
\subfigure{(b) $2$nd moment}
\label{fig:min-relative-uncertainty-2nd-moment}
\end{minipage} \\
\begin{minipage}[t]{0.5\hsize}
\centering
\includegraphics[clip,width=0.8\textwidth]
{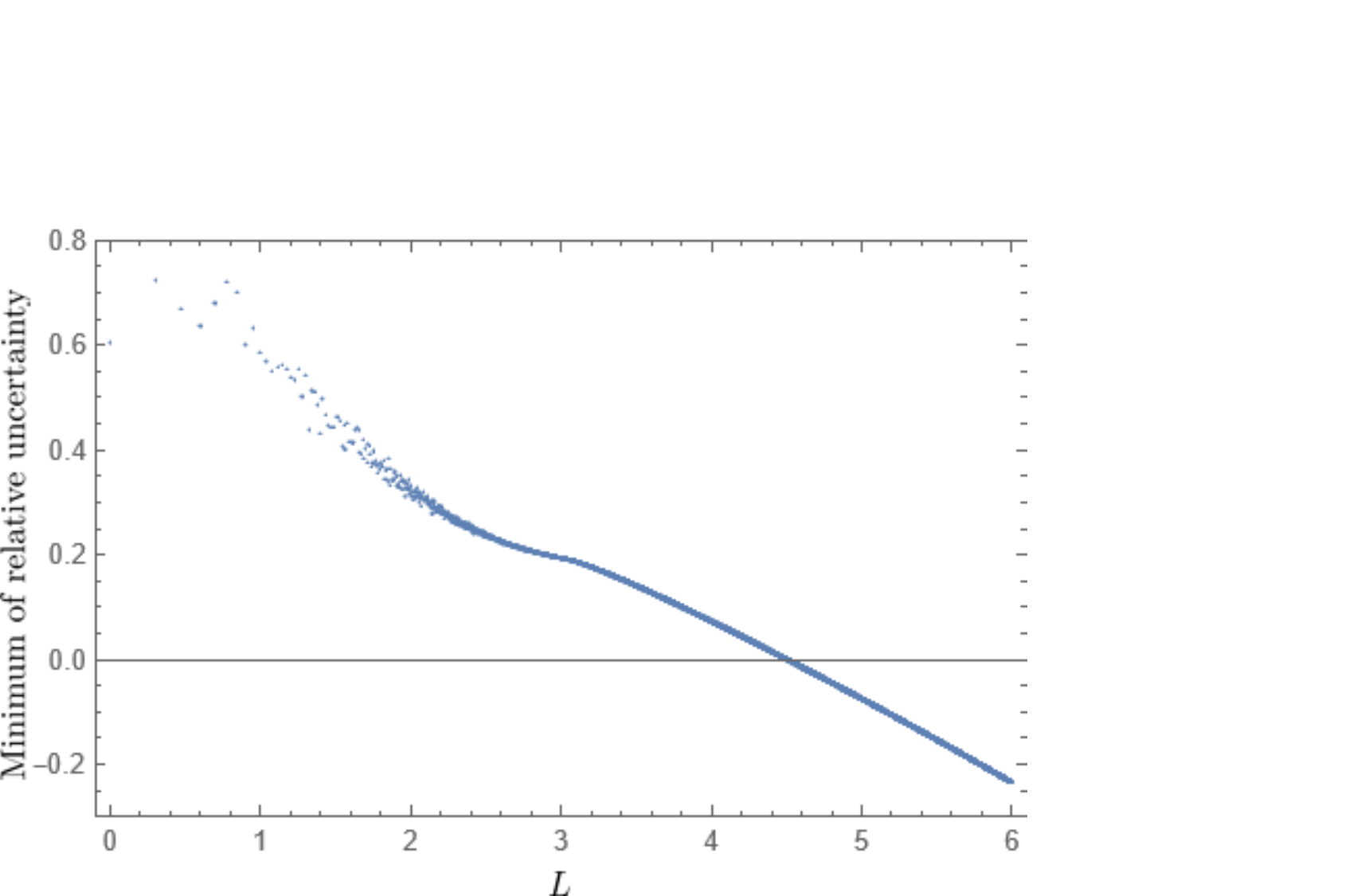}
\subfigure{(c) $3$rd moment }
\label{fig:min-relative-uncertainty-3rd-moment}
\end{minipage} &
\begin{minipage}[t]{0.5\hsize}
\centering
\includegraphics[clip,width=0.8\textwidth]
{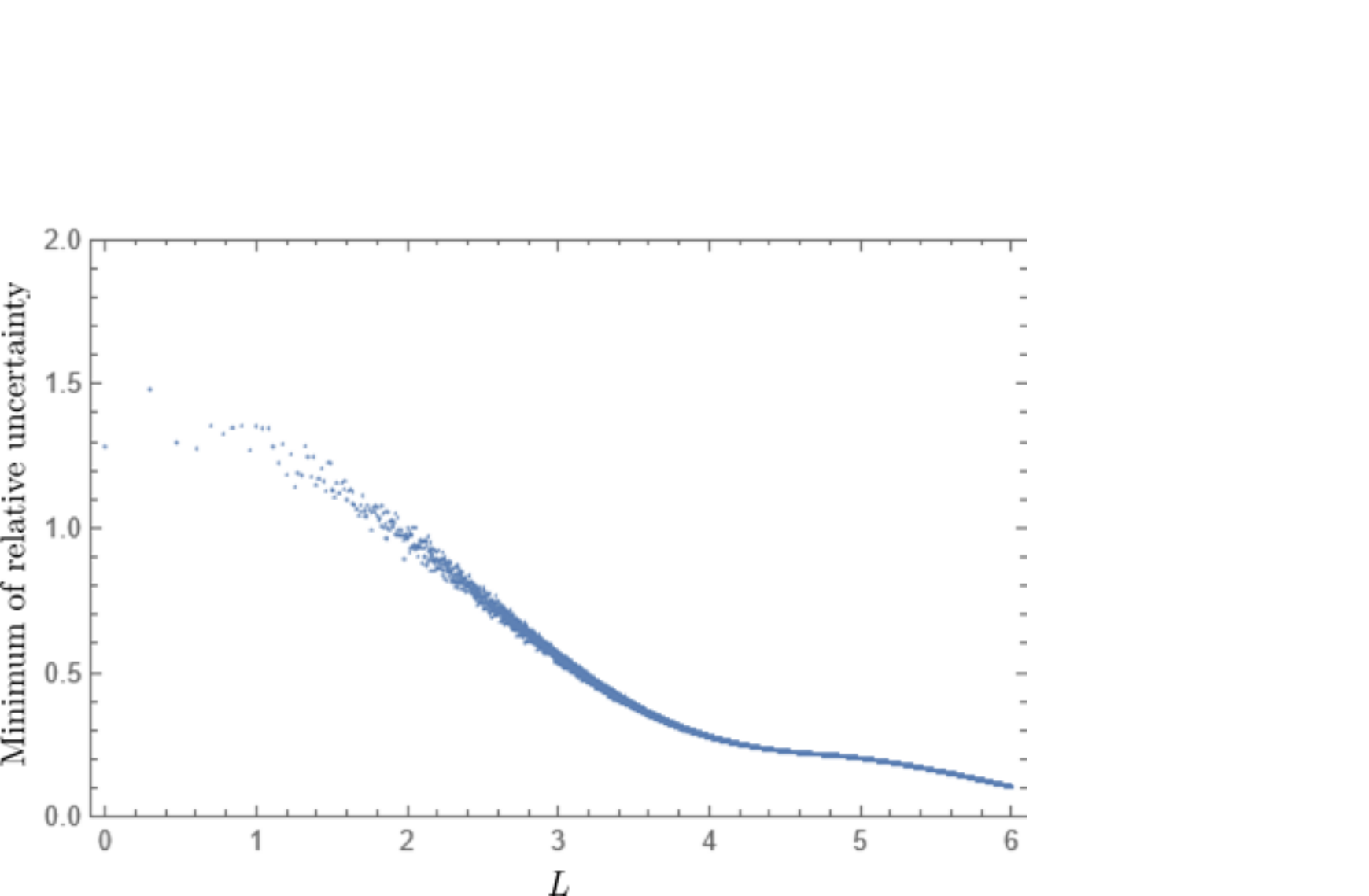}
\subfigure{ (d) $4$th moment }
\label{fig:min-relative-uncertainty-4th-moment}
\end{minipage}
\end{tabular}
\caption{
The log-log plot of the minimum relative uncertainty of each moment.
}
\label{fig:minimum-relative-uncertainty-at-L}
\end{figure*}
\subsubsection{Anonymity}
We consider a case that Eve performs some POVM measurement
on the state at the end of step 4 to estimate the value of $\omega_{l}$, 
and we evaluate the upper bound of the information gained by Eve. 
For this purpose, 
we calculate quantum Fisher information (QFI) about $\omega_{l}$.
After the interaction with the magnetic field at the step 4, 
the quantum Fisher information does not change as long as we perform unitary operators that are independent of $\omega_{l}$. 
So, to obtain the upper bound, 
we can calculate the QFI of the state at the beginning of the step 7. 
From Eq.\eqref{eq:phi_4_ket} that is the definition of $\ket{\phi_4}$,
we define $\ket{\phi_4}$ which belongs to the $j$-th copy system as follows.
\begin{eqnarray}
\ket{\phi_4(t)}_{j}
\equiv
\frac{1}{\sqrt{L}}
\sum_{l=0}^{L-1}
\ket{l}_{\rm{r},\textit{j}} \ket{+_{\omega_l t } }_{\rm{d},\textit{j}} 
\end{eqnarray}
where the definition of $\ket{+_{\theta} }$ is 
Eq.\eqref{eq:+_theta_ket} and we define 
$
\hat{\rho}_{fin,j}(t)
\equiv
\ket{\phi_4(t)}_j \bra{\phi_4(t)}
$
.
Then, 
the state at the beginning of the step 7 is given as follows.
\begin{eqnarray}
{}
&{}&
\hat{ \rho}_{fin,all}
(\Delta t ,k)
 \nonumber \\
&\equiv&
\hat{\rho}_{fin,0} (0)
\otimes 
\hat{\rho}_{fin,1} ( \Delta t)
\otimes 
 \nonumber \\
&\cdots&
\otimes 
\hat{\rho}_{fin,k-1} ( (k-1) \Delta t)
\otimes 
\hat{\rho}_{fin,k} ( k \Delta t)
 \nonumber \\
\end{eqnarray}
where we traced out the sensor qubits because there is 
no entanglement between the sensor qubits and the other qubits.
We consider to estimate $\omega_l$
at each place $l \in V$, and so
we calculate
the symmetric logarithmic derivative (SLD) as 
quantum Fisher information (QFI) 
for multiparameter estimation \cite{Liu_2020}.
We consider parameters $a,b$ for state $\hat{\rho}$.
SLD is defined as operator $\hat{L}_a$ which satisfies the following relationship.
\begin{eqnarray}
\frac
{ \partial \hat{\rho}}
{ \partial a}
=
\frac{1}{2}
( \hat{\rho} \hat{L}_{a} + \hat{L}_{a} \hat{\rho} )
\end{eqnarray}
Then, QFI can be rewritten as
\begin{eqnarray}
F_{a,b}
=
Tr
\biggl(
\hat{L}_b 
\frac
{ \partial \hat{\rho}}
{ \partial a}
\biggr)
.
\end{eqnarray}
Therefore, for the state
$
\hat{ \rho}_{fin,all}
(\Delta t ,k)
$,
we can calculate the QFI as follows.
\begin{eqnarray}
F_{l,l^{\prime}}
=
\frac
{ 2L \delta_{l,l^{\prime}} -1 }{L^2} 
(\Delta t)^2
\sum_{j=1}^{k}
j^2
\end{eqnarray}
We also calculate the inverse matrix of the QFI, 
and the matrix elements are given as follows.
\begin{eqnarray}
(F^{-1})_{l,l^{\prime}}
=
\frac
{1 + L \delta_{l,l^{\prime}}}
{2 (\Delta t)^2
\sum_{j=1}^{k} j^2}
\end{eqnarray}
The quantum Cramer-Rao bound is calculated as follows.
\begin{eqnarray}
( \Delta \omega_{l}^{est} )^{2}
&\ge&
\frac{(F^{-1})_{l,l}}{N} 
=
\frac
{L+1}
{2N (\Delta t)^2
\sum_{j=1}^{k} j^2}
 \nonumber \\
\end{eqnarray}
So the lower bound of the uncertainty is given as follows.
\begin{eqnarray}
\Delta \omega_{l}^{est} 
&\ge&
\frac{1}{\Delta t}
\sqrt{
\frac
{L+1}
{2N \sum_{j=1}^{k} j^2}
}
\end{eqnarray}
Since the minimum interaction time with the magnetic fields is $\Delta t$, 
the uncertainty of the phase with the minimum interaction time is given as follows.
\begin{eqnarray}
\Delta \theta_{l}^{est} 
\equiv
\Delta \omega_{l}^{est} \cdot \Delta t
&\ge&
\sqrt{
\frac
{L+1}
{2N \sum_{j=1}^{k} j^2}
}
\end{eqnarray}
As the uncertainty becomes larger, 
it becomes more difficult to know the precise value of the magnetic field. 
So, in the limit of large $L$, 
the security is guaranteed in the sense that no one can obtain the information of the individual magnetic field from the quantum state.
On the other hand, when we consider a finite $L$, 
the uncertainty is also finite.
In our paper, 
we define that our protocol is secured if the following condition is satisfied.
\begin{eqnarray}
\Delta \theta_{l}^{est} 
&\ge&
\sqrt{
\frac
{L+1}
{2N \sum_{j=1}^{k} j^2}
}
\ge
\pi
\end{eqnarray}
\subsection{Evaluation of relative uncertainty of moments}
We evaluate the performance of our protocol by using numerical simulation.
We randomly generate the magnetic fields $ \{\omega_{l}\}_{l=0}^{L-1}$ from 
a uniform distribution where 
the maximum value $\omega_{max}$ is 5 and the minimum value $\omega_{min}$ is 1.
In numerical simulations, 
we set parameters where the condition 
$\sqrt{
\frac
{L+1}
{2N \sum_{j=1}^{k} j^2}
}
=
\pi$ is satisfied.
We plot the relative uncertainty of estimation for the
$k$-th ($k=1,2,3,4$) order moment in Figure 
\ref{fig:relative-uncertainty-of-moment}.
Here,
the horizontal axis denotes the time difference $\Delta t$.
Also, 
the vertical axis denotes the logarithms of relative uncertainties in 
Eqs. 
\eqref{eq:relative_uncertainty_at_even_order_moment}
and 
\eqref{eq:relative_uncertainty_at_odd_order_moment}
with a base of $10$.
The phase of quantum states should be less than $2\pi$.
Therefore the maximim time difference is as follows
\begin{eqnarray}
\label{eq:maximim_time_difference}
\Delta t_{max}
&\equiv&
\frac
{ 2\pi }
{ \omega_{max} k}
\end{eqnarray}
where $\omega_{max}$ ($k$) is the maximum value of magnetic fields 
(order of moment).
Due to the relationship of $L \propto N$, 
we can suppress the relative uncertainty as we increase $L$.
Also, as we increase the order of the moment, 
the relative uncertainty becomes larger.
Comparing different orders of moment,
the smaller order of moment is,
the smaller relative uncertainty of the moment is.
Note that from 
Eq. \eqref{eq:maximim_time_difference}, 
$\omega_{max} = 5$ and $2\pi \approx 6.28 $,
$ \Delta t < 1$ is satisfied when $2 \le k$.
So, as we increase
$k$, 
$ (\Delta t)^{k} $ becomes smaller, which increases
the statistical error $ \delta (C^{2}) (t=0, \Delta t, k) $.
If we decrease $\Delta t$, 
the systematic error becomes smaller while the statistical error becomes larger.
Due to this tradeoff relationship,
there exists an optimal $\Delta t$ to minimize the uncertainty.
When we consider the first (second) order moment,
the relative uncertainty becomes $1\%$
for $L=10^8$ ($L=1.7 \times 10^9$)
from Fig.
\ref{fig:relative-uncertainty-of-moment} (a) ( (e) )
.
We plot the minimum relative uncertainty of the $k$-th order moment in Fig.\ref{fig:minimum-relative-uncertainty-at-L}.
Here, the horizontal axis denotes logarithm of $L$ 
and the vertical axis denotes the minimum relative uncertainty
$\underset{\Delta t}{\text{min }}\omega_{\text{relative uncertainty}} 
(\Delta t, k=m )$.
\section{Conclusion}
In conclusion, 
we have proposed an anonymous estimation of the intensity distribution 
of magnetic fields with a quantum sensor network.
Suppose that there are magnetic fields in $L$ different places.
The purpose of our protocol is to estimate statistical quantities 
(such as average and variance) of the target fields 
at different places without knowing the individual value of the target fields.
There are sensor qubits, register qubits, and a data qubit.
Firstly,
the distributer prepares a separable initial state 
and sends this to the controller.
Secondly,
the controller performs a CSWAP operation between them,
and sends all sensor qubits to each sensor holder.
Thirdly,
each sensor holder lets the sensor qubit interact 
with a local magnetic field and 
sends the state back to the controller.
Finally,
the controller locally performs the CSWAP operation and
performs measurements.
From the measurement results, 
we can estimate the values of the statistical quantities.
Importantly, in the limit of a large $L$, it becomes impossible to extract information of the local magnetic fields from the state in our protocol.
On the other hand, in our protocol, 
we can estimate theoretically any moment of the field distribution by measuring a specific observable and 
evaluated relative uncertainty of $k$-th ($k=1,2,3,4$) order moment.
Since the magnetic-field sensor is used in medical
science and material engineering, 
our protocol could play an important role to protect confidential information 
once the quantum network becomes available.
\par
This work was supported by MEXT KAKENHI (Grant Nos. 20H05661).
This work was also supported 
by JST Moonshot R\&D (Grant Number JPMJMS226C). 
Y. Takeuchi is supported by the MEXT Quantum Leap Flagship Program (MEXT Q-LEAP) Grant Number JPMXS0120319794, JST [Moonshot R\&D -- MILLENNIA Program] Grant Number JPMJMS2061, and the Grant-in-Aid for Scientific Research (A) No.JP22H00522 of JSPS.
\section{Appendix}
\begin{widetext}
\subsection{Schematic of our protocol by using a single system without copies}
In the main text, we consider $(k+1)$ copies. 
On the other hand, to get an intuition, 
we illustrate our protocol when we use a single system
without copies.
\label{sec:analysis_of_protocol_flow}
\begin{enumerate}
\item The distributer sends qubits of the initial state to the controller. \\

The initial state $\ket{\phi_0}$ is composed of 
register qubits $\ket{+^{ log_2 L}}_{\rm{r}}$
, a data qubit $\ket{+}_{\rm{d}}$
, and sensor qubits $\ket{0^L}_{\rm{s}}$ as follows.
\begin{eqnarray}
\ket{\phi_0}
&\equiv&
\ket{+^{log_2 L}}_{\rm{r}} \ket{+}_{\rm{d}} \ket{0^L}_{\rm{s}}
\nonumber \\
&=&
\frac{1}{ \sqrt{ 2^{ {log_2 L} } } }
\sum_{l=0}^{L-1}
\ket{l}_{\rm{r}} \ket{+}_{\rm{d}} \ket{0^{L}}_{\rm{s}}
\nonumber \\
&=&
\frac{1}{\sqrt{L} }
\sum_{l=0}^{L-1}
\ket{l}_{\rm{r}} \ket{+}_{\rm{d}} \ket{0^{L}}_{\rm{s}}
\end{eqnarray}
where the number of register (sensor) qubits is $\log _2 L$ ($L$). 
Then, 
the distributer sends this initial state 
to the controller. 
Here, subscripts of each ket,
$\rm{r}$, $\rm{d}$ and $\rm{s}$ denote register qubit, 
data qubit, and sensor qubit, respectively.
\item The controller performs the CSWAP between the sensor qubits and the data qubit.\\

We define the CSWAP operation as follows.
\begin{eqnarray}
\hat{U}_{\rm{CSWAP}}
&=&
\sum_{l=0}^{L-1}
\ket{l}_{\rm{r}} \bra{l}
\otimes
\hat{U}_{\rm{SWAP}}(\rm{d},\textit{l})
\end{eqnarray}
Here, 
$\hat{U}_{\rm{SWAP}}(\rm{d},\textit{l})$ denotes a SWAP operation between the data qubit and the $l$-th sensor qubit.
\begin{eqnarray}
\ket{\phi_0}
\xrightarrow[CSWAP]{}
\ket{\phi_1}
\equiv
\frac{1}{\sqrt{L}}
\sum_{l=0}^{L-1}
\ket{l}_{\rm{r}} \ket{0}_{\rm{d}} \ket{+}_{l} 
\ket{0^{(L-1) }}_{\rm{s} \setminus \{ \textit{l} \}}
\end{eqnarray}
where the subscript $l$ denotes the $l$-th sensor qubit 
and $\rm{s} \setminus \{ l\}$ denotes all sensor qubits 
except for the $l$-th sensor qubit.
\item For $l=0,1,\cdots , L-1$, 
the controller sends the $l$-th sensor qubit to the $l$-th sensor holder.\\

\item Each sensor holder lets his/her sensor 
qubit interacts with the magnetic fields for a time $t$.\\

The unitary operator induced by the magnetic fields is described as follows.
We assume that the magnetic fields are applied along the $z$ axis, 
and the amplitude of the magnetic fields at the site $l$ is $\omega_l$.
The total Hamiltonian of the magnetic fields is described as follows.
\begin{eqnarray}
\hat{H}
\equiv
\frac{1}{2}
\sum_{l=0}^{L-1}
\omega_l \hat{\sigma}_{z}^{(l)}
\end{eqnarray}
The unitary evolution induced by this Hamiltonian is described as follows.
\begin{eqnarray}
\hat{U}_{B}(t)
\equiv
e^{ -i\hat{H}t }
=
e^{ -\frac{it}{2} \sum_{l=0}^{L-1} \omega_{l} \hat{\sigma}_{z}^{(l)} }
=
\bigotimes_{l=0}^{L-1}
e^{ -\frac{1}{2} i \omega_{l} t \hat{\sigma}_{z}^{(l)} }
\end{eqnarray} 
The state after the time evolution is as follows.
\begin{eqnarray}
\ket{\phi_2(t)}
&\equiv&
\hat{U}_{B}(t)
\ket{\phi_1}
\nonumber \\
&=&
\biggl(
\prod_{j =0 }^{L-1}
e^{- \frac{ i \omega_j t }{2} }
\biggr)
\frac{1}{\sqrt{L}}
\sum_{l=0}^{L-1}
\ket{l}_{\rm{r}} \ket{0}_{\rm{d}} 
\ket{+_{\omega_l t } }_{l} 
\ket{0^{ (L-1) }}_{\rm{s} \setminus \{ \textit{l} \} }
\end{eqnarray}
where we use the following notation 
\begin{eqnarray}
\label{eq:+_theta_ket}
\ket{+_{\theta} }
\equiv
\frac{1}{\sqrt{2}}
( \ket{0}+ e^{i \theta} \ket{1} ) 
.
\end{eqnarray}
We can remove the global phase, and we obtain
\begin{eqnarray}
\ket{\phi_2(t)}
\equiv
\hat{U}_{B}(t)
\ket{\phi_1}
=
\frac{1}{\sqrt{L}}
\sum_{l=0}^{L-1}
\ket{l}_{\rm{r}} \ket{0}_{\rm{d}} 
\ket{+_{\omega_l t } }_{l} 
\ket{0^{ (L-1) }}_{\rm{s} \setminus \{ \textit{l} \} }
.
\end{eqnarray}
\item The $l$-th sensor holder sends the sensor 
qubit back to the controller.\\

\item The controller performs $\hat{U}_{\rm{CSWAP}}$.\\
\begin{eqnarray}
\ket{\phi_2(t)}
\xrightarrow[CSWAP]{}
\ket{\phi_3(t)}
\equiv
\frac{1}{\sqrt{L}}
\sum_{l=0}^{L-1}
\ket{l}_{\rm{r}} \ket{+_{\omega_l t } }_{\rm{d}} 
\ket{0^{ L }}_{\rm{s}}
\end{eqnarray}
There is no entanglement between the sensor qubits
and the other qubits,
and so we can trace out the sensor qubits. We obtain.
\begin{eqnarray}
\label{eq:phi_4_ket}
\ket{\phi_4(t)}
\equiv
\frac{1}{\sqrt{L}}
\sum_{l=0}^{L-1}
\ket{l}_{\rm{r}} \ket{+_{\omega_l t } }_{\rm{d}} 
\end{eqnarray}
where we have 
$\ket{\phi_3(t)}
=
\ket{\phi_4(t)} \otimes \ket{0^{ L }}_{\rm{s}}$
.
\item The controller performs a POVM measurement on register qubits and a data qubit.\\
We choose the following POVM.
\begin{eqnarray}
\begin{cases}
M_{\psi, 1}
=
\ket{+^{ log_2 L}}_{\rm{r}} \bra{+^{ log_2 L}} 
\otimes
\ket{\psi}_{\rm{d}} \bra{\psi}
\\
M_{\psi, 2}
=
I^{ \otimes log_2 L}_{\rm{r}} \otimes I_{\rm{d}} 
-
M_{\psi, 1}
\end{cases}
\end{eqnarray}
Let us define this POVM as POVM($\ket{\psi}$).
Let us consider a state when the POVM $M_{\psi, 1}$ occurs.
\begin{eqnarray}
M_{\psi, 1}
\ket{\phi_4}
&=&
\ket{+^{ log_2 L}}_{\rm{r}} 
\otimes
\ket{\psi}_{\rm{d}} 
\braket{\psi|A }
\end{eqnarray}
where
\begin{eqnarray}
\label{eq:Aket}
\ket{A}
&\equiv&
\frac{1}{L}
\sum_{l=0}^{L-1}
\ket{+_{\omega_l t} }
=
\frac{1}{\sqrt{2}}
( \ket{0} + B(t) \ket{1} )
.
\end{eqnarray}
Then, the probability is calculated as follows.
\begin{eqnarray}
P_{1}(\psi)
&\equiv&
\|
M_{\psi, 1}
\ket{\phi_4}
\|^2
=
| \braket{\psi|A } |^2
\end{eqnarray}
This probability does not change 
when we swap $\omega_l$ with $\omega _{l'}$ for any $l$ and $l'$.
This means that the measurement result does not contain 
any information about where the magnetic field $\omega_l$
is generated.
The state after this POVM is as follows.
\begin{eqnarray}
\ket{\phi ( \psi ) }
&\equiv&
\frac
{ M_{\psi, 1} \ket{\phi_4} }
{ \| M_{\psi, 1} \ket{\phi_4} \| }
=
\frac
{ M_{\psi, 1} \ket{\phi_4} }
{ \sqrt{ P_{1}(\psi) } }
=
\frac
{ M_{\psi, 1} \ket{\phi_4} }
{ | \braket{\psi|A } | }
=
\frac
{ \braket{\psi|A } }
{ | \braket{\psi|A } | }
\ket{+^{ log_2 L}}_{\rm{r}} 
\otimes
\ket{\psi}_{\rm{d}} 
\end{eqnarray}
Let us consider the case that 
POVM $M_{\psi, 2} = I - M_{\psi, 1}$ occurs. 
In this case, the probability 
$P_{2}(\psi)=1-P_{1}(\psi)$ is also symmetric.
However,
since
the identity $I$ reproduces the original asymmetric state,
the state after the POVM $M_{\psi, 2}$ is not symmetric.
So, if Eve steals the quantum state, 
it is in principle possible to extract the information about where the magnetic fields $\omega_l$ are generated.
\item Repeat the above steps $N$ times.\\
\end{enumerate}
Let us define eigenstates of $ \hat{\sigma}_{x} (\hat{\sigma}_{y}) $ 
as $\ket{\pm,x}$ ($\ket{\pm,y}$).
The probability that $M_{\ket{\pm,x}, 1}$
($M_{\ket{\pm,y}, 1}$) occurs is defined as $P_{\pm,x}$ ($P_{\pm,y}$). 
We can calculate these probabilities as follows.
\begin{eqnarray}
\label{eq:probability}
\begin{cases}
P_{\pm,x}
\equiv
\bigl\|
\ket{\pm ,x } \bra{\pm ,x } \cdot \ket{A}
\bigr\|^{2}
=
\bigl|
\braket{\pm ,x | A}
\bigr|^{2}
=
\frac{1}{ 4 }
\{ 1 + | B(t) |^{2} \pm 2 Re(B(t)) \}
\\
P_{\pm,y}
\equiv
\bigl\|
\ket{\pm ,y } \bra{\pm ,y } \cdot \ket{A}
\bigr\|^{2}
=
\bigl|
\braket{\pm ,y | A}
\bigr|^{2}
=
\frac{1}{ 4 }
\{ 1 + | B(t) |^{2} \pm 2 Im(B(t)) \}
\end{cases}
\end{eqnarray}
The real part and imaginary part of $B(t)$ are
represented as follows.
\begin{eqnarray}
\begin{cases}
Re(B(t)) 
=
P_{+,x} - P_{-,x} 
=
\braket{A | \hat{\sigma}_{x} | A}
\\
Im(B(t)) 
=
P_{+,y} - P_{-,y} 
=
\braket{A | \hat{\sigma}_{y} | A}
\end{cases}
\end{eqnarray}
\subsection{Combined measurement}
\label{sec:overall_measurement}
Let us consider a state $\hat{\rho} (t)$. 
By changing the variable $t$,
we can consider a composite state as follows.
\begin{eqnarray}
\hat{ \tilde{\rho}} (t, \Delta t ,k )
&\equiv&
\hat{\rho}_{0} (t) \otimes \hat{\rho}_{1} (t+ \Delta t ) \otimes
\cdots
\otimes \hat{\rho}_{k-1} (t+ (k-1) \Delta t ) \otimes \hat{\rho}_{k} (t+ k \Delta t )
\nonumber \\
&=&
\bigotimes_{j=0}^{k}
\hat{\rho}_{j} (t+ j \Delta t )
\end{eqnarray}
where $\hat{\rho}_j$ denotes the $j$-th copy of 
the state by changing the variable $t$.
When we measure an observable $\hat{A}$,
the expectation is described as follows.
\begin{eqnarray}
A(t)
\equiv
\braket{\hat{A}}_{E}
&=&
\mathrm{Tr} 
( \hat{\rho} (t) \hat{A} )
\end{eqnarray}
As an observable for the composite system, 
we consider the following.
\begin{eqnarray}
\hat{C}( \Delta t,k)
&\equiv&
\frac{1}
{ ( \Delta t)^k }
\sum_{j=0}^{k}
(-1)^{k+j}
\binom{k}{j}
\hat{A}_{j}
\end{eqnarray}
$\hat{A}_{j}$ denotes an observable on the $j$-th copy,
and this is defined as follows.
\begin{eqnarray}
\hat{A}_{j}
&\equiv&
\hat{I}_{0}
\otimes
\hat{I}_{1}
\otimes
\cdots
\otimes
\hat{I}_{j-1}
\otimes
\hat{A}_{j}
\otimes
\hat{I}_{j+1}
\otimes
\cdots
\otimes
\hat{I}_{k-1}
\otimes
\hat{I}_{k}
\end{eqnarray}
By measuring $\hat{C}( \Delta t,k)$, 
we obtain the $k$-th order forward difference of
the expectation of physical quantity $\hat{A}$.
\begin{eqnarray}
C(t,\Delta t, k)
\equiv
\braket{ \hat{C}( \Delta t, k) }_{E}
&=&
\mathrm{Tr} 
( \hat{ \tilde{\rho}} (t, \Delta t ,k ) 
\hat{C}( \Delta t, k) )
\nonumber \\
&=&
\mathrm{Tr} 
\biggl\{
\biggl( 
\bigotimes_{j=0}^{k}
 \hat{\rho}_{j} (t+ j \Delta t ) 
 \biggr) 
\biggl(
\frac{1}
{ ( \Delta t)^k }
\sum_{l=0}^{k}
(-1)^{k+l}
\binom{k}{l}
\hat{A}_{l} 
 \biggr)
 \biggr\}
\nonumber \\
&=&
\frac{1}
{ ( \Delta t)^k }
\sum_{l=0}^{k}
(-1)^{k+l}
\binom{k}{l}
\mathrm{Tr} 
\biggl\{
\biggl( 
\bigotimes_{ \substack {
j=0 \\j \neq l }
}^{k}
 \hat{\rho}_{j} (t+ j \Delta t ) 
 \biggr) 
 \otimes
\biggl(
 \hat{\rho}_{l} (t+ l \Delta t ) 
\hat{A}_{l} 
 \biggr)
 \biggr\}
\nonumber \\
&=&
\frac{1}
{ ( \Delta t)^k }
\sum_{l=0}^{k}
(-1)^{k+l}
\binom{k}{l}
\mathrm{Tr} 
\biggl(
 \hat{\rho}_{l} (t+ l \Delta t ) 
\hat{A}_{l} 
 \biggr)
 \cdot 
\prod_{ \substack {
j=0 \\j \neq l }
}^{k}
\mathrm{Tr} 
\biggl( 
 \hat{\rho}_{j} (t+ j \Delta t ) 
 \biggr) 
\nonumber \\
&=&
\frac{1}
{ ( \Delta t)^k }
\sum_{l=0}^{k}
(-1)^{k+l}
\binom{k}{l}
 A (t+ l \Delta t ) 
\nonumber \\
&=&
\frac
{ \Delta^k A(t)}
{ \Delta t^k }
\end{eqnarray}
Since the forward difference is not exactly the same as the differentiation, 
we define such a deviation as $\epsilon (t, \Delta t ,k)$ as follows.
\begin{eqnarray}
\frac
{ \Delta^k A(t)}
{ \Delta t^k }
=
\frac
{ d^k A(t)}
{ d t^k }
+
\epsilon
(t, \Delta t ,k)
\end{eqnarray}
As long as the number of measurements is finite, 
we cannot perfectly estimate the expected value of
$\hat{C}( \Delta t, k)$. 
This means that we cannot estimate the forward difference 
due to the shot noise.
Assuming that the number of measurements is finite,
we define our estimate of the forward difference as
\begin{eqnarray}
\frac
{ \Delta^k A^{est}(t)}
{ \Delta t^k }
.
\end{eqnarray}
Up to now, 
we consider to prepare the state and perform the POVM measurement $N$ times. 
However, 
this is equivalent to prepare $N$ copies 
of the state and measure only one time, 
and we adopt this strategy to make the calculation simple. 
The larger composit system is described as follows.
\begin{eqnarray}
\hat{ \rho}^{ \prime } (t, \Delta t ,k ,N)
&\equiv&
\hat{ \tilde{\rho}}^{(1)} (t, \Delta t ,k )
\otimes 
\hat{ \tilde{\rho}}^{(2)} (t, \Delta t ,k )
\otimes
\cdots
\otimes
\hat{ \tilde{\rho}}^{(N-1)} (t, \Delta t ,k )
\otimes 
\hat{ \tilde{\rho}}^{(N)} (t, \Delta t ,k )
\nonumber \\
&=&
\bigotimes_{l=1}^{N}
\hat{ \tilde{\rho}}^{(l)} (t, \Delta t ,k )
\end{eqnarray}
$ \hat{ \tilde{\rho}}^{(l)} (t, \Delta t ,k )$ is
the $l$-th copy of the composit system.
Then, we consider the following observable for the larger system.
\begin{eqnarray}
\hat{D}( \Delta t, k, N)
&\equiv&
\frac{1}{ N }
\sum_{l=1}^{N}
\hat{C}^{(l)} ( \Delta t, k)
\end{eqnarray}
$ \hat{C}^{(l)} $ denotes an observable on
the $l$-th copy of the composite system. 
This is described as follows.
\begin{eqnarray}
\hat{C}^{(l)} ( \Delta t, k)
&\equiv&
\hat{I}^{(1)}
\otimes
\hat{I}^{(2)}
\otimes
\cdots
\otimes
\hat{I}^{(l-1)}
\otimes
\hat{C}^{(l)} ( \Delta t, k)
\otimes
\hat{I}^{(l+1)}
\otimes
\cdots
\otimes
\hat{I}^{(N-1)}
\otimes
\hat{I}^{(N)}
\end{eqnarray}
We can calculate the expectation value of this observable as follows.
\begin{eqnarray}
D(t,\Delta t, k, N)
&\equiv&
\braket{ \hat{D}(\Delta t, k, N) }_{E}
\nonumber \\
&=&
\mathrm{Tr} 
( 
\hat{ \rho}^{ \prime } (t, \Delta t ,k ,N)
\hat{D}( \Delta t, k, N)
)
\nonumber \\
&=&
\mathrm{Tr} 
\biggl( 
\hat{ \tilde{\rho}} (t, \Delta t ,k )
\hat{C}( \Delta t, k)
\biggr)
=
\braket{ \hat{C}(\Delta t, k) }_{E}
=
C(t,\Delta t, k)
=
\frac
{ \Delta^k A(t)}
{ \Delta t^k }
\end{eqnarray}
We consider the variance of $ \hat{D}( \Delta t, k, N) $ and
$ \hat{C}( \Delta t, k) $.
The square of $ \hat{C}(\Delta t, k) $ is calculated as follows.
\begin{eqnarray}
\{ \hat{C} ( \Delta t, k) \}^{2}
&=&
\frac{1}{ ( \Delta t)^{2k} }
\biggl\{
\sum_{l=0}^{k}
 \binom{k}{l}^{2}
\hat{A}_{l}^{2}
+
2
\sum_{\substack{ 
l,m=0 \\
l < m
} }^{k}
(-1)^{l+m}
\binom{k}{l}
\binom{k}{m}
\hat{A}_{l} 
\hat{A}_{m} 
\biggr\}
\end{eqnarray}
The expectation of the square of $ \hat{C}(\Delta t, k) $ is 
calculated as follows.
\begin{eqnarray}
{}
&{}&
( C^{2} ) (t,\Delta t, k)
\equiv
\braket{ ( \hat{C}^{2} ) (\Delta t, k) }_{E}
=
\mathrm{Tr} 
[
\hat{ \tilde{\rho}} (t, \Delta t ,k ) 
\{ \hat{C}( \Delta t, k) \}^{2}
]
\nonumber \\
&=&
\frac{1}{ ( \Delta t)^{2k} }
\sum_{l=0}^{k}
 \binom{k}{l}^{2}
(A^{2}) (t+ l \Delta t ) 
+
\frac{2}{ ( \Delta t)^{2k} }
\sum_{\substack{ 
l,m=0 \\
l < m
} }^{k}
(-1)^{l+m}
\binom{k}{l}
\binom{k}{m}
A (t+ l \Delta t ) 
A (t+ m \Delta t ) 
\nonumber \\
\end{eqnarray}
The square of the expected value of $ \hat{C}(\Delta t, k) $ is as follows.
\begin{eqnarray}
{}
&{}&
\{C(t,\Delta t, k)\}^{2}
\nonumber \\
&=&
\biggl\{
\frac{1}
{ ( \Delta t)^k }
\sum_{l=0}^{k}
(-1)^{k+l}
\binom{k}{l}
A (t+ l \Delta t ) 
\biggr\}^{2}
\nonumber \\
&=&
\frac{1}{ ( \Delta t)^{2k } }
\sum_{l=0}^{k}
\binom{k}{l}^{2}
\{ A (t+ l \Delta t ) \}^{2}
+
\frac{2}{ ( \Delta t)^{2k } }
\sum_{\substack{ 
l,m=0 \\
l < m
} }^{k}
(-1)^{l+m}
\binom{k}{l}
\binom{k}{m}
A (t+ l \Delta t ) 
A (t+ m \Delta t ) 
\nonumber \\
\end{eqnarray}
$ \delta (C^{2}) (t,\Delta t, k) $ 
(the variance of $ \hat{C}(\Delta t, k)$ ) is 
calculated as follows.
\begin{eqnarray}
{}
&{}&
\delta (C^{2}) (t,\Delta t, k) 
\equiv
( C^{2} ) (t,\Delta t, k)
-
\{C(t,\Delta t, k)\}^{2}
\nonumber \\
&=&
\frac{1}{ ( \Delta t)^{2k} }
\sum_{l=0}^{k}
 \binom{k}{l}^{2}
[
\mathrm{Tr} 
\{
\hat{\rho} (t+ l \Delta t ) 
\hat{A}^{2} 
\}
-
\bigl(
\mathrm{Tr} 
(
\hat{\rho} (t+ l \Delta t ) 
\hat{A}
)
\bigr)^{2} 
]
\end{eqnarray}
By considering the central limit theorem,
$ \delta (D^{2}) (t,\Delta t, k, N)$
(the variance of $\hat{D} ( \Delta t, k, N )$ ) 
is calculated as follows.
\begin{eqnarray}
\delta (D^{2}) (t,\Delta t, k, N)
&\equiv&
( D^{2} ) (t,\Delta t, k, N)
-
\{ D(t,\Delta t, k, N) \}^{2}
=
\frac{
\delta (C^{2} ) (t,\Delta t, k)
}{N}
\end{eqnarray}
In the limit of a large number of measurements, 
we can perfectly estimate the value of 
the $k$-th order forward difference, 
and so we have
\begin{eqnarray}
\lim_{N \to \infty}	
\frac
{ \Delta^k A^{est}(t)}
{ \Delta t^k }
=
\frac
{ \Delta^k A(t)}
{ \Delta t^k }
\end{eqnarray}
The variance of the estimation of forward difference under 
the finite number of measurements is calculated as follows.
\begin{eqnarray}
\delta (D^{2}) (t,\Delta t, k, N)
&\equiv&
\biggl<
\biggl( 
\frac
{ \Delta^k A^{est}(t)}
{ \Delta t^k }
-
\frac
{ \Delta^k A(t)}
{ \Delta t^k }
\biggr)^2
\biggr>_{E}
\nonumber \\
&=&
\biggl<
\biggl(
\frac
{ \Delta^k A^{est}(t)}
{ \Delta t^k }
-
\frac
{ d^k A(t)}
{ d t^k }
-
\epsilon
(t, \Delta t ,k)
\biggr)^2
\biggr>_{E}
\nonumber \\
&=&
\biggl<
\biggl(
\frac
{ \Delta^k A^{est}(t)}
{ \Delta t^k }
-
\frac
{ d^k A(t)}
{ d t^k }
\biggr)^2
-
2
\biggl(
\frac
{ \Delta^k A^{est}(t)}
{ \Delta t^k }
-
\frac
{ d^k A(t)}
{ d t^k }
\biggr)
\epsilon
(t, \Delta t ,k)
+
\epsilon
(t, \Delta t ,k)^2
\biggr>_{E}
\nonumber \\
&=&
\biggl<
\biggl(
\frac
{ \Delta^k A^{est}(t)}
{ \Delta t^k }
-
\frac
{ d^k A(t)}
{ d t^k }
\biggr)^2
\biggr>_{E}
-
2
\biggl<
\biggl(
\frac
{ \Delta^k A^{est}(t)}
{ \Delta t^k }
-
\frac
{ d^k A(t)}
{ d t^k }
\biggr)
\biggr>_{E}
\epsilon
(t, \Delta t ,k)
+
\epsilon
(t, \Delta t ,k)^2
\nonumber \\
&=&
\biggl<
\biggl(
\frac
{ \Delta^k A^{est}(t)}
{ \Delta t^k }
-
\frac
{ d^k A(t)}
{ d t^k }
\biggr)^2
\biggr>_{E}
-
\epsilon
(t, \Delta t ,k)^2
\end{eqnarray}
So we obtain the following.
\begin{eqnarray}
\biggl<
\biggl(
\frac
{ \Delta^k A^{est}(t)}
{ \Delta t^k }
-
\frac
{ d^k A(t)}
{ d t^k }
\biggr)^2
\biggr>_{E}
&=&
\delta (D^{2}) (t,\Delta t, k, N)
+
\epsilon
(t, \Delta t ,k)^2
=
\frac{
\delta (C^{2}) (t,\Delta t, k) 
}{N}
+
\epsilon
(t, \Delta t ,k)^2
\nonumber \\
\end{eqnarray}
\end{widetext}
\bibliographystyle{apsrev4-1}
 \bibliography{bibref}
\end{document}